\documentclass[a4paper,11pt]{article}
\pdfoutput=1
\usepackage{jcappub}
\usepackage{amssymb,amsmath,mathrsfs,enumerate}
\usepackage{graphicx,rotate,multicol}
\usepackage{float}

\usepackage{subfig}

\usepackage{slashed}
\usepackage{mathtools}
\usepackage{multirow}

\allowdisplaybreaks

\title{\boldmath Gravitational Wave Probe of Primordial Black Hole Origin via Superradiance}
\author{Indra Kumar Banerjee,}
\author{Ujjal Kumar Dey}
\affiliation{Department of Physical Sciences, Indian Institute of Science Education and Research Berhampur,\\Transit Campus, Government ITI, Berhampur 760010, Odisha, India}

\emailAdd{indrab@iiserbpr.ac.in}
\emailAdd{ujjal@iiserbpr.ac.in}

\abstract{In this article we have used stochastic gravitational wave background as a unique probe to gain insight regarding the creation mechanism of primordial black holes. We have considered the cumulative gravitational wave background which consists of the primary part coming from the creation mechanism of the primordial black holes and the secondary part coming from the different mechanisms the primordial black holes go through. We have shown that in the presence of light or ultra light scalar bosons, superradiant instability generates the secondary part of the gravitational wave background which is the most detectable. In order to show the unique features of the cumulative background, we have consdiered the delayed vacuum decay during a first order phase transition as the origin of primordial black holes. We have shown the dependence of the features of the cumulative background, such as the mass of the relevant light scalars, peak frequencies, etc. on the transition parameters. We have also generated the cumulative background for a few benchmark cases to further illustrate our claim.}

\begin{document}
\maketitle
\flushbottom

%~~~~Keywords: 

%%%%%%%%%%%%%%%%%%%%%%%%%%%%%%%%%%%%%%%%%%%%%%%%%%%%%%
%%%%%%%%%%%%%%%%%%%%%%%%%%%%%%%%%%%%%%%%%%%%%%%%%%%%%%
\section{Introduction}
\label{sec:intro}
\begin{figure}[H]
\centering
\includegraphics[scale=0.65]{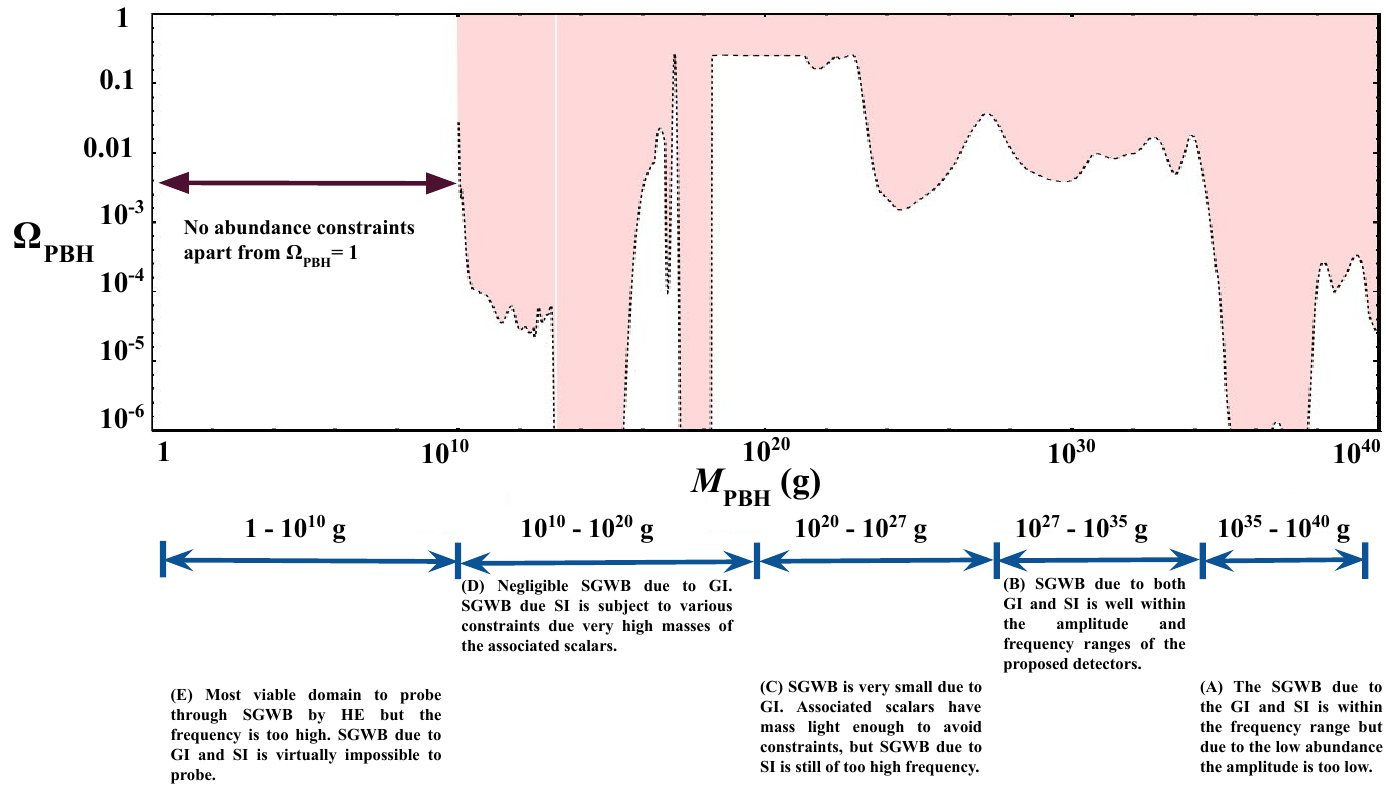}
\caption{This cartoon shows the approximate constraints on the density of PBHs $(\Omega_{\mathrm{PBH}})$ and the possibility of SGWB from the PBHs through GI, HE and SI. It is to be noted here that all PBHs below the mass of $10^{16}\mathrm{~g}$ has already evaporated.}
\label{cartoon}
\end{figure}

%%%%%%%%%%%%%%%%%%%%%%%%%%%%%%%%%%%%%%%%%%%%%%%%%%%%%%
%%%%%%%%%%%%%%%%%%%%%%%%%%%%%%%%%%%%%%%%%%%%%%%%%%%%%%
One of the biggest unanswered questions in modern physics is the nature of dark matter (DM). Some proposed theories suggest that DM could be particulate, e.g., weakly interacting massive particles, axions, and ALPs. In contrast, some theories suggest that some primordial exotic or compact objects, such as primordial black holes (PBH), can account for some or all of the DM.
%
%Standard model of particle physics and standard cosmology, though tremendously successful in explaining most of the universe, still fail to answer some of the questions. One of those questions are the origin and the nature of dark matter. Dark matter in principle could be explained by different fields (e.g. axions, ALPs) or it could be some compact object of primordial origin, like primordial black hole (PBH). 
%
Primordial back holes were first proposed in \cite{Zeldovich:1967lct}, and since then, it has been the focus of the community. The LIGO and Virgo collaborations~\cite{LIGOScientific:2016aoc} discovered gravitational waves (GW) almost a decade ago, which took the attention once again towards PBHs. There exist many proposed mechanisms regarding the formation of the PBHs in literature, e.g., collapse from inhomogeneities during radiation-~\cite{Carr:1974nx,Grillo:1980rt} and matter-dominated era~\cite{Khlopov:1980mg, Khlopov:1985jw,Carr:1994ar}, critical collapse~\cite{Niemeyer:1997mt, Niemeyer:1999ak}, collapse in single~\cite{Carr:1993aq, Bullock:1996at, Saito:2008em} and multi-field inflationary models~\cite{Randall:1995dj, Garcia-Bellido:2016dkw, Braglia:2020eai}, collapse of cosmic string loops~\cite{Hawking:1987bn, Borah:2023iqo}, collapse during a first order phase transition (FOPT)~\cite{Crawford:1982yz} to name a few. In the past few years, studies of novel model-dependent~\cite{Kawana:2021tde, Baker:2021nyl,Huang:2022him}, and model-independent~\cite{Kawana:2022olo, Liu:2021svg,Gouttenoire:2023naa,Lewicki:2023ioy} mechanisms regarding formation of PBH during FOPTs have appeared. Among these, the model-independent mechanisms discuss the creation of PBHs during FOPT due to the collapse of overdense regions formed due to the asynchronous bubble nucleation. 
Furthermore, many of these PBH forming mechanisms, i.e., scalar perturbation~\cite{Matarrese:1993zf, Matarrese:1996pp, Matarrese:1997ay}, inflation~\cite{Khlebnikov:1997di, Easther:2006vd, Easther:2007vj,Choudhury:2013woa}, cosmic strings~\cite{Vilenkin:1981bx, Vachaspati:1984gt, Hindmarsh:1994re}, FOPT~\cite{Witten:1984rs, Hogan:1986qda} also serves as sources of stochastic gravitational wave background (SGWB) in the Universe.
This SGWB and the properties of the PBH, i.e., their mass and abundance, depend on the mechanisms through which they were created. Apart from this, PBHs, if they exist, may go through many exciting processes, such as gravitational interaction (GI) among each other and with other astrophysical black holes (ABH), they may evaporate through Hawking evaporation (HE)~\cite{Hawking:1974rv} into standard model (SM) or beyond the standard model (BSM) degrees of freedom~\cite{Ireland:2023avg}, they may go through superradiant instability (SI) if specific masses of scalar bosons exist~\cite{Yang:2023aak,Tsukada:2018mbp,Berti:2019wnn,Arvanitaki:2010sy}, etc. In principle, these mechanisms can also give rise to SGWB in the range of detection proposed by the detectors depending on a few factors, i.e., the mass and the abundance of the PBH, the mass of the BSM scalar bosons (if they exist). In Fig.~\ref{cartoon} we have expressed the possibility of generating detectable SGWB from PBHs through these mechanisms for a vast range of PBH masses and their abundances.
Therefore, we claim that both the SGWB directly generated by the cosmological events responsible for the creation of PBH (scalar perturbations, cosmic strings, FOPT, inflation, etc.) and the SGWB, which are later created by these PBHs are also an indirect result of these events. In our previous work, we have already given this proposal, though we had specifically chosen the gravitational interactions in that case. Hence, the unique properties related to those creation mechanisms are encoded in the features of these \textit{cumulative SGWB}. Following the footsteps of our previous work, we take FOPTs as the creation mechanism and illustrate this further. The mass and abundance of the PBHs born during a FOPT is the consequence of the properties of the FOPT, i.e., the transition strength ($\alpha$), duration of the FOPT ($1/\beta$), the energy content of the universe during the transition, critical temperature ($T_{\mathrm{cri}}$), etc. Similarly, the SGWB generated due to the FOPT depends on these parameters. Therefore, the combined spectra, i.e., the sum of the spectrum arising from the FOPT and the spectrum arising from the mechanism related to the PBHs, are a result of the FOPT, where the former is a direct consequence, and the latter is an indirect consequence of the FOPT.

The literature on probing the PBHs utilizing the features in the gravitational waves is quite large~\cite{Borah:2023iqo, Gehrman:2023esa, Franciolini:2022htd, Acuna:2023bkm, Gehrman:2022imk, Mandic:2016lcn, Chen:2018rzo, Sugiyama:2020roc, Garcia-Bellido:2021jlq, Cui:2021hlu, Papanikolaou:2022chm,Xie:2023cwi,Barman:2022pdo, Agashe:2022jgk,Datta:2023xpr,Wang:2016ana,Wang:2019kaf,Wang:2021djr,Choudhury:2023rks,Choudhury:2023jlt}\footnote{It is to be noted that in recent time there have been studies to probe PBH using methods other than GW as well~\cite{Basilakos:2023xof,Carr:2023tpt,Karmakar:2023hlb,Saha:2021pqf,Arimoto:2021cwc}.}. However, our previous study was one of the first ones to conduct a study that uses the cumulative GW spectra~\cite{Banerjee:2023brn}. In our previous work, we have shown the secondary spectra only due to the gravitational interactions of the PBHs among themselves and among other ABHs. This secondary spectra is robust because since PBHs of those mass range exist and they are indeed as abundant as we had considered them to be, then the secondary spectra is a direct consequence of general relaivity. However, any new physics (Hawking evaporation, superradiant instability) was not considered, which, if present, can give rise to much larger secondary spectra. In this article, we conduct a more general study by considering other generation mechanisms for the indirect SGWB spectra, i.e. SGWB due to superradiant instability in presence of ultra light bosons (ULBs)\footnote{Though vector and tensor bosons can also create superradiant instability, in this work we only consider scalar ULBs.}, to probe FOPT as the origin of PBH. It is worth mentioning that many BSM theories give rise to ULBs, such as (pseudo-)scalar particles in axiverse, light gauge bosons etc~\cite{Goodsell:2009xc,Holdom:1985ag}. Recent studies have shown the effect of axiverse on PBH evaporation and superradiance~\cite{Calza:2023rjt}. Our study also provides a new framework to use this axiverse in probing the existence and other features of PBH. From the perspective of particle physics, these ULBs can be characterized by their mass and their couplings (if any) with other SM fields. If PBHs exist, then the existence (non-existence) of this kind of a secondary spectra can favour 
(disfavour) certain mass ranges of these ULBs. It is also to be noted that ULBs can have mass anywhere between $10^{-33}\mathrm{~eV}$ to $1\mathrm{~eV}$~\cite{Ferreira:2020fam} which in turn provides a huge mass range of PBHs which can go through superradiant instability. Within this range, ULBs with mass above $10^{-22}\mathrm{~eV}$ can play the role of dark matter. In principle, a study of this nature can be performed for any mechanism that generates PBH and SGWB. It is to be noted that in most cases, the SGWB is created due to the PBH having an amplitude such that it can only be detected in the frequency range $\mathcal{O}(10^{-9})\mathrm{~Hz}$ to $\mathcal{O}(10^3)\mathrm{~Hz}$. In order to investigate the PBHs for which this creation mechanism serves as the origin of the most detectable SGWB spectra, we constrain ourselves in the domain where PBH masses are between $\mathcal{O}(10^{-4})M_{\odot}$ and $\mathcal{O}(10^2)M_{\odot}$. This is because the gravitational interactions and superradiant instability give rise to very high frequency GW for PBHs with masses lower than this, which is outside the detection range. In contrast, for masses higher than this, the amplitude of the GW is lower (due to the abundance constraints) than the detection range. It is to be mentioned that the GW due to the Hawking evaporation is ultra high frequency for any PBH with mass over $\mathcal{O}(1)\mathrm{~g}$~\cite{Ireland:2023avg}. Therefore, we have refrained from considering them in this study. In this study, we consider FOPTs, which produce PBHs in the mass range mentioned above, and obtain and discuss the properties of the cumulative SGWB due to those FOPTs.
This article is organized as follows: in Sec. \ref{sec:fopt}, we discuss FOPT as the common origin of SGWB and the PBHs. After discussing the mass and abundance of PBHs formed out of FOPT, we go on to show the SGWB spectrum that can originate from the FOPT. In Sec. \ref{sec:supradgw}, we discuss the SGWB spectrum originating mainly from the superradiant instability. We present our results in Sec. \ref{sec:results} and summarize and conclude in Sec. \ref{sec:concl}.

%%%%%%%%%%%%%%%%%%%%%%%%%%%%%%%%%%%%%%%%%%%%%%%%%%%%%%
%%%%%%%%%%%%%%%%%%%%%%%%%%%%%%%%%%%%%%%%%%%%%%%%%%%%%%
\section{FOPT as the Origin of the PBH and GW}
\label{sec:fopt}
%%%%%%%%%%%%%%%%%%%%%%%%%%%%%%%%%%%%%%%%%%%%%%%%%%%%%%
%%%%%%%%%%%%%%%%%%%%%%%%%%%%%%%%%%%%%%%%%%%%%%%%%%%%%%
Throughout the literature, different mechanisms have been proposed regarding the creation mechanism of PBH. Our study considers the mechanisms proposed in ~\cite{Kawana:2022olo,Liu:2021svg,Gouttenoire:2023naa,Lewicki:2023ioy}.

The universe might have gone through many phase transitions (PT) during its evolution as the temperature of the universe decreased. These phase transitions could be of first or second order, depending on the true and false minima distribution. The temperature at which the phase transition begins, i.e., at which the pre-existing minima becomes a local minima and a global minima is created, is called the critical temperature ($T_{\mathrm{cri}}$). If a potential barrier exists between these two minima, then the transition is of the first order. In these situations, the universe goes to the true vacuum from the false vacuum through thermal or quantum tunnelling. Physically, this happens through the nucleation of bubble of true vacuum, which is a probabilistic process. As the temperature reduces further, there exists a temperature where the probability of the nucleation of a bubble in a Hubble region is unity. This is called as the nucleation temperature ($T_{\mathrm{nuc}}$).

Since this bubble nucleation is a probabilistic process, there might be situations where, in some Hubble regions, the bubble nucleation is delayed. In that case, the Hubble region is dominated by the vacuum energy, and as the universe expands, the vacuum energy density remains constant, whereas the radiation energy density is diluted. In a while, if there is still no bubble nucleation in the abovementioned Hubble region, then the vacuum energy is converted to radiation energy density, and an overdense region is created. If the overdensity of the overdensity is above a specific threshold value, then the region collapses into a PBH. This mechanism is susceptible to the FOPT parameters, i.e., the transition has to have enough duration for the overdense region to occur. Therefore, in this mechanism, $\beta/H$ is taken to be $\leq\mathcal{O}(10)$. The FOPT has to be strong enough so that enough energy is released and the overdensity can form, and therefore $\alpha$ is taken to be $\geq\mathcal{O}(1)$. It is worth mentioning that the basic mechanism considered in Refs.~\cite{Kawana:2022olo,Liu:2021svg,Gouttenoire:2023naa,Lewicki:2023ioy} is the same as the one mentioned above. However, they differ in a few subtleties and, therefore, predict slightly different values of PBH abundances for the same values of $\alpha$ and $\beta/H$. We remain agnostic about those details and take a few benchmark points for the FOPTs we consider in this work, which are motivated by different BSM scenarios~\cite{Freese:2023fcr,Carena:2019une} and are given in Tab. \ref{table_FOPT}.

\begin{table}[H]
\centering
\begin{tabular}{|l|l|l|l|l|}
\hline
\multicolumn{1}{|c|}{Parameters} & \multicolumn{1}{c|}{I} & \multicolumn{1}{c|}{II} & \multicolumn{1}{c|}{III}\\ \hline \hline
$T$ (in GeV) & 10  & 1 & 0.039\\ \hline
$\alpha$    & 2.135                   & 1.2 	&1\\ \hline
$\beta/H$   & 16                  & 10	& 8\\ \hline
$\kappa$   & 0.5                  & 1 	& 1\\ \hline
\end{tabular}
\caption{The relevant parameters describing the different FOPTs.}
\label{table_FOPT}
\end{table}
Here $\kappa$ is the fraction of vacuum energy transferred into the bubble walls. Now we will use these values and the expressions mentioned above to obtain the mass and the abundance of the PBHs and the gravitational waves due to collision of the bubble walls.
%%%%%%%%%%%%%%%%%%%%%%%%%%%%%%%%%%%%%%%%%%%%%%%%%%%%%%
\subsection{Creation of PBHs: Mass and Abundance}
\label{cre_pbh}
%%%%%%%%%%%%%%%%%%%%%%%%%%%%%%%%%%%%%%%%%%%%%%%%%%%%%%
In the previous paragraph we briefly mentioned the mechanism of PBH formation during the FOPT. The PBH formed during and FOPT depends only on the transition temperature, which can be expressed as,
\begin{align}
M(T)=\gamma\frac{4\pi}{3}\bar{\rho}(T)H^{-3}(T),
\label{pbh_mass}
\end{align}
where, $H^2(t) = \bar{\rho}(t)/3M^2_{\text{P}},$
with $\bar{\rho}$ is the average energy density and $M_{\text{P}}$ is the Planck mass. Here $\gamma$ ($\le 1$) is a numerical factor which depends on the specifics of the gravitational collapse. 
%Now, we assume that most of the PBHs which are formed during a FOPT have a mass, $M_{\mathrm{PBH}}=M(t_{1.45})$, where $t_{1.45}$ is defined by,
%
%\begin{align}
%\frac{\rho_{\mathrm{inside}}(t_{1.45})}{\rho_{\mathrm{outside}}(t_{1.45})}%=1.45.
%\end{align}
%
%In order to obtain $t_{1.45}$ one needs to simultaneously solve Eqs \ref{friedmann_eq} and \ref{den_evol}. Then we put, $t_{1.45}$ in Eq. \ref{pbh_mass} to obtain the PBH mass.
%
%Using this prescription, the PBH masses we get for the three different FOPT given in Tab. \ref{table_FOPT} are, $50~M_{\odot}$, $0.5~M_{\odot}$ and $3\times 10^{-5}~M_{\odot}$ respectively.%
%
%In order to obtain the abundance of the PBH today, we first calculate the transition probability,
%\begin{align}
%p(t)=-\dfrac{dP(t)}{dt}.
%\label{trans_prob}
%\end{align}
%The fraction of the PBHs when they were formed are given by~\cite{Kawana:2022olo},
%begin{align}
%B(M)=\dfrac{MH^3(t)}{\pi T}\dfrac{p(t)}{s(T)}\left(\dfrac{d\ln M}{dt}\right)^{-1},
%\end{align}
%where, $s(T)$ is the entropy density of the plasma at temperature $T$ and $p(t)$ is the transition probability of an overdense region to collapse into a PBH.
%Finally, the abundance of the PBH, i.e. the fraction of dark matter which is made of PBH can be obtained by,
%\begin{align}
%f_{\mathrm{PBH}}(M)=\dfrac{1}{\Omega_{\mathrm{DM}}}\left(\dfrac{M_{\mathrm{eq}}}{M}\right)^{1/2}B(M),
%\end{align}
%where $\Omega_{\mathrm{DM}}=0.264$~\cite{Planck:2018vyg} is the density parameter of dark matter and $M_{\mathrm{eq}}=2.8\times  10^{17}~M_{\odot}$~\cite{Kawana:2022olo}.

All the Ref. mentioned above follows he same consideration regarding the PBH mass, however they differ in their arguments of the PBH abundance as is it is very sensitive transition temperature, strength and duration. In this regard, we follow the formalism of Ref.~\cite{Liu:2021svg} and obtain the mass of PBH and their abundances corresponding to the benchmark points (BP) characterized by the parameters given in Tab.~\ref{table_FOPT} are calculated to be $\{4.6\times10^{-4}M_{\odot}$, $0.01\}$, $\{4.6\times10^{-2}M_{\odot}$, $0.02\}$ and $\{30M_{\odot}$, $0.006\}$ for BP I, BP II, and BP III,  respectively. It is worth mentioning that all these BPs gives rise PBHs which are within the abundance constraints given by Subaru HSC~\cite{Niikura:2017zjd}, Kepler~\cite{Griest:2013esa}, OGLE~\cite{Niikura:2019kqi}, MACHO/EROS~\cite{Macho:2000nvd,EROS-2:2006ryy}, SNe~\cite{Zumalacarregui:2017qqd}, Ly-$\alpha$~\cite{Murgia:2019duy}, and CMB~\cite{Poulin:2017bwe}.
%The mass of the PBHs and their abundances along with the relevant constraints have been depicted in Fig.~\ref{mass_abun}.
%
%\begin{figure}[H]
%\centering
%\includegraphics[scale=0.5]{Abundance}
%\caption{Abundances for different PBH masses. The constraints from Subaru HSC~\cite{Niikura:2017zjd}, Kepler~\cite{Griest:2013esa}, OGLE~\cite{Niikura:2019kqi}, MACHO/EROS~\cite{Macho:2000nvd,EROS-2:2006ryy}, SNe~\cite{Zumalacarregui:2017qqd}, Ly-$\alpha$~\cite{Murgia:2019duy}, CMB~\cite{Poulin:2017bwe} have also been shown.}
%\label{mass_abun}
%\end{figure}

%%%%%%%%%%%%%%%%%%%%%%%%%%%%%%%%%%%%%%%%%%%%%%%%%%%%%%
\subsection{GW Background from FOPT}
\label{gw_fopt}
\begin{figure}[t]
\centering
\includegraphics[scale=0.4]{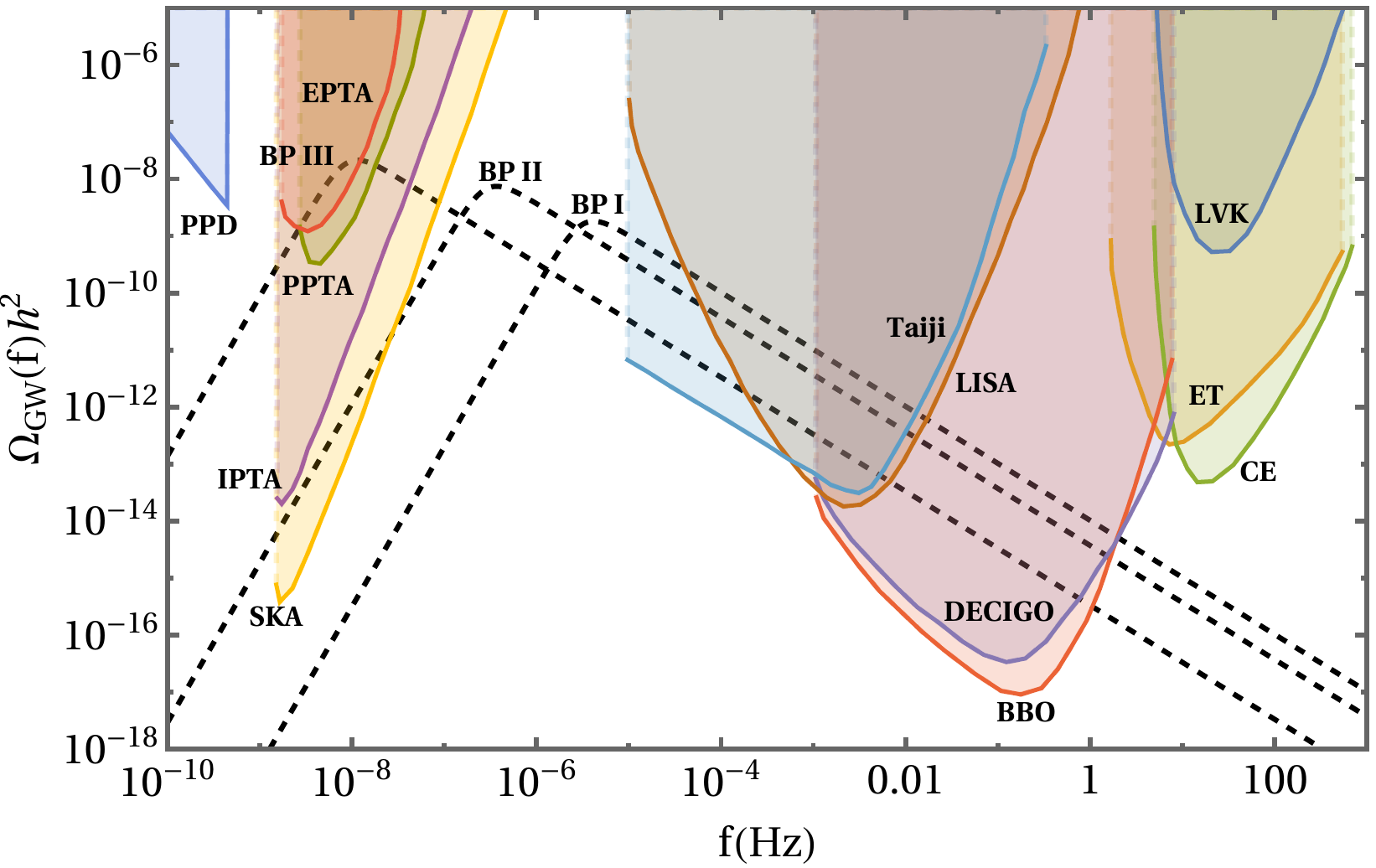}
\caption{The SGWB spectrum for the two first order phase transitions considered in this paper. The sensitivity curves for SKA~\cite{Carilli:2004nx}, IPTA~\cite{Hobbs:2009yy}, EPTA~\cite{Lentati:2015qwp}, PPTA~\cite{Shannon:2015ect}, Taiji~\cite{Ruan:2018tsw}, LISA~\cite{LISA:2017pwj}, DECIGO~\cite{Kawamura:2011zz}, BBO~\cite{Phinney:2004bbo}, CE~\cite{Reitze:2019iox}, ET~\cite{Punturo:2010zz}, LIGO, Virgo and KAGRA (LVK)~\cite{Somiya:2011np,LIGOScientific:2014pky} have been shown.}
\label{fopt_sgwb}
\end{figure}

%%%%%%%%%%%%%%%%%%%%%%%%%%%%%%%%%%%%%%%%%%%%%%%%%%%%%%
The expression for the SGWB spectrum due to FOPT can be expressed as~\cite{Kosowsky:1992vn},
\begin{align}
\Omega_{\mathrm{GW}}(f)&=\dfrac{1}{\rho_c}\dfrac{d\rho_{\mathrm{GW}}}{d\ln f} \nonumber \\
&=1.67\times 10^{-5}\left(\dfrac{H}{\beta}\right)^2 \left(\dfrac{\kappa\alpha}{1+\alpha}\right)^2 \dfrac{0.11v_w^2}{0.42+v_w^3}\left(\dfrac{100}{g_{*}}\right)^{1/3} \dfrac{3.8(f/f_p)^{2.8}}{1+2.8(f/f_p)^{3.8}},
\label{FOPT_GW}
\end{align}
where,
\begin{align}
f_p=\dfrac{0.62}{1.8-0.1v_w+v_w^2}\left(\dfrac{\beta}{H}\right)\dfrac{T}{100\mathrm{~GeV}}\left(\dfrac{g_{*}}{100}\right)^{1/6}\times 1.65\times 10^{-5} \mathrm{~Hz},
\label{FOPT_fp}
\end{align}
is the peak frequency. In this study, we only consider the contributions due to the bubble collisions as they are the significant contributors to the SGWB. It is also worth mentioning that since we do not consider supercooled PT, the transition, nucleation, and percolation temperature are of the same order, and we consider that all the quantities related to the FOPT are evaluated at $T_{\mathrm{cri}}$. Now, using the BP values in Tab. \ref{table_FOPT} in Eqs.~\eqref{FOPT_GW} and \eqref{FOPT_fp}, we get the SGWB spectrum which has been presented in Fig. \ref{fopt_sgwb}.\footnote{Apart from pulsar timing arrays and laser interferometer experiments, which are sensitive to GW signals between $10^{-9}-10^{3}\mathrm{~Hz}$, other experimental methods, such as, mechanical resonators, GW-electromagnetic wave conversion have been proposed which can in principle probe GW signal upto $10^{18}\mathrm{~Hz}$~\cite{Franciolini:2022htd}. However, their proposed sensitivity is much lower than the domain of our interest and hence we do not include them.}
%
%%%%%%%%%%%%%%%%%%%%%%%%%%%%%%%%%%%%%%%%%%%%%%%%%%%%%%
%%%%%%%%%%%%%%%%%%%%%%%%%%%%%%%%%%%%%%%%%%%%%%%%%%%%%%
\section{Generation of Secondary GW}
\label{sec:supradgw}
%%%%%%%%%%%%%%%%%%%%%%%%%%%%%%%%%%%%%%%%%%%%%%%%%%%%%%
%%%%%%%%%%%%%%%%%%%%%%%%%%%%%%%%%%%%%%%%%%%%%%%%%%%%%%
As mentioned in the previous section, cosmological FOPTs create PBHs and SGWB spectra. These PBHs then might go through different processes to create other SGWB, which we term \textit{secondary GW spectra} as they are, in most cases, much less in its amplitude than the one created during FOPT. The main mechanisms that can create these secondary spectra are (i) gravitational interactions between the PBHs among themselves and other astrophysical black holes (ABH), (ii) evaporation of these PBHs into graviton, and (iii) superradiant instability of the PBHs due to the presence of ultralight bosons (ULB). It is to be noted that the three creation mechanisms have zero, one (Hawking evaporation) and two (superradiance and ULBs) new physics, respectively. Now these different mechanisms are most effective for PBHs in different mass ranges, e.g., gravitational interactions and superradiant instabilities due to ULBs creating detectable secondary GW for PBH masses over $\mathcal{O}(10)M_{\odot}$ but for the same mass range hawking evaporation create very high frequency GW ($\mathcal{O}(10^{32}$ Hz)) which is well beyond the scope of any proposed GW observation experiments~\cite{Ireland:2023avg}.

In principle, PBHs created during cosmological FOPTs can have masses from $\mathcal{O}(1)\mathrm{~g}$ to $\mathcal{O}(1000)M_{\odot}$, but any PBH of mass less than $10^{16}\mathrm{~g}$ have already evaporated. In this article, we constrain ourselves to PBHs which are not already evaporated and can create secondary GWs detectable through running or proposed GW observation experiments. Therefore, we contain our parameter space to PBHs of mass in the range $(10^{-6}M_{\odot},10^3M_{\odot})$ as in this domain, the abundance of the PBHs can be high enough for the GW signals to be in the detectable range. We now break this domain into two parts, \textit{mid-mass} PBHs which have mass range from $10^{-6}M_{\odot}$ to $1M_{\odot}$ and \textit{high-mass} which have mass range from $1M_{\odot}$ to $10^3M_{\odot}$. This breakdown has been done because mid-mass PBHs can only create appreciable secondary GW through superradiant instability, whereas high-mass PBHs can create secondary GW through gravitational interaction and superradiant instability.

In order to have superradiant instability, the PBHs have to have a non-vanishing spin, which can trigger it. The PBH then creates the ULB cloud around itself at the cost of its angular momentum when the Compton wavelength of the ULB becomes comparable to the PBH size, and hence, it slows down. This cloud of ULBs create a ``gravitational atom" in which, analogous to the electron in the hydrogen atom, ULBs have discrete energy states. This PBH-ULB system (gravitational atom) which then can generate GW through different mechanisms, among which we focus on the transition of energy levels of the atom. In this regard, the gravitational fine structure constant is defined as $\alpha_g = GM_{\mathrm{PBH}}m_{a}/\hslash c$ where $m_{a}$ is the mass of the ULB, $G$ is the Newton's constant, $\hslash$ is the Planck's constant and $c$ is the speed of light in vacuum. In this study, we consider the transition process from the excited state of $(n,l,m)$ from $(3,1,1)$ to $(2,1,1)$. If $\alpha_g\leq 0.5$, then the above transition dominates, and energy is radiated as GWs. The energy radiated in the form of gravitational radiation can be expressed as~\cite{Yang:2023aak},
\begin{align}
E_{\mathrm{GW}}=\dfrac{\alpha M_{\mathrm{PBH}}}{3m_{a}}\left(1-e^{-t/\tau_{\mathrm{GW}}}\right)\Delta\omega,
\end{align}
where
\begin{align}
\Delta\omega&=\dfrac{5}{72}\alpha_g^{2}m_{a} \\ 
\tau_{\mathrm{GW}}&=1.23\times 10^{34}\left(\dfrac{m_{a}}{10^{-29}\mathrm{~GeV}}\right)^{-9}\left(\dfrac{M_{\mathrm{PBH}}}{10^{6}M_{\odot}}\right)^{-8}\mathrm{~yr}.
\end{align}
Here, $\Delta\omega$ is the frequency split between the two states mentioned above, $\tau_{\mathrm{GW}}$ is the timescale of the gravitational radiation. It is to be noted here, that due to the limit on $\alpha_g$ only a small range of mass of the ULBs contribute significantly to the gravitational radiation for a specific mass of a PBH.

Generally, two scenarios are taken for the GW generation due to superradiant instability: (i) superradiant instability from spinning isolated PBHs and (ii) superradiant instabilities from the remnant of the mergers of two PBHs. Since we consider the creation of the PBH due to FOPT during the readiation dominated phase of the universe, the initial spin of the PBHs is vanishingly small. However, the merger between two equal mass PBHs with vanishing spin creates a remnant with diemnsionless spin parameter spin $\chi\sim 0.68$ and final mass of that of twice the initial mass of the progenitor PBH~\cite{Barausse:2009uz}. Therefore, though it has been shown that isolated PBHs will lead to GW spectra that are much higher in amplitude than that of the merger remnants to remain physically accurate, we consider the latter in this work.

The SGWB spectra due to the superradiant instability of the merger remenats can be epressed as~\cite{Yang:2023aak,Garcia-Bellido:2021jlq},
\begin{align}
\Omega_{\mathrm{GW}}^{\mathrm{sup}}=\dfrac{f}{\rho_c}\int dz\dfrac{dt}{dz}\int \tau p(\chi)d\chi\dfrac{dE_s}{df_s},
\end{align} 
where
\begin{align}
\tau &= \int \dfrac{dm_1}{m_1}\dfrac{dm_2}{m_2}\dfrac{d\tau}{d\mathrm{ln}m_1 d\mathrm{ln}m_2},\\
\dfrac{d\tau}{d\mathrm{ln}m_1 d\mathrm{ln}m_2} &= 14.8 \mathrm{yr}^{-1}\mathrm{Gpc}^{-3}f(m_1)f(m_2)\dfrac{(m_1 + m_2)^{10/7}}{(m_1m_2)^{5/7}} \left(\dfrac{\delta_{\mathrm{loc}}}{10^8}\right)\left(\dfrac{v_0}{10\mathrm{km}/\mathrm{s}}\right)^{-11/7},\\
\dfrac{dt}{dz} &= \dfrac{1}{H_0\sqrt{\Omega_{\mathrm{M}}(1+z)^3+\Omega_{\Lambda}}(1+z)},\\
\dfrac{dE_s}{df_s} &= E_{\mathrm{GW}}\delta\left(f(1+z)-f_s\right).
\end{align}
Here $\tau$ is the number of mergers per unit time per comoving volume, $\delta_{\mathrm{loc}}$ is the local density contrast, $v_{0}$ is the relative initial velocity between two of the progenitor PBHs, $dt/dz$ is the rate of change of the look back time with respect to the redshift, $H_0$ is the present day Hubble parameter, $\rho_c$ is the critical density of the universe, $\Omega_{\mathrm{M}}$ and $\Omega_{\Lambda}$ are the matter density and the dark energy density of the universe and $f_s$ is the frequency at the source frame. For these scenarios, the source produces the gravitational radiation at a constant frequency, which can be expressed as $f_s = m_a c^2/\pi \hslash$. Here, $p(\chi)$ is the distribution function of the PBH spin. Since we only consider merger remnants, we take $p(\chi)=\delta (\chi -0.68)$~\cite{Barausse:2009uz}.

Apart from this, gravitational interactions, such the formation of binary and the eventual merger and close hyperbolic encounters of PBHs can also create secondary GW spectra. The mechanism and the required definitions are provided in Ref.~\cite{Banerjee:2023brn}. However, for our parameter space those GW are negligible as explained in Sec. \ref{highpbh}.
%%%%%%%%%%%%%%%%%%%%%%%%%%%%%%%%%%%%%%%%%%%%%%%%%%%%%%
\subsection{Mid-mass PBH}
\label{midpbh}
\begin{figure}[t]
\centering
\includegraphics[scale=0.5]{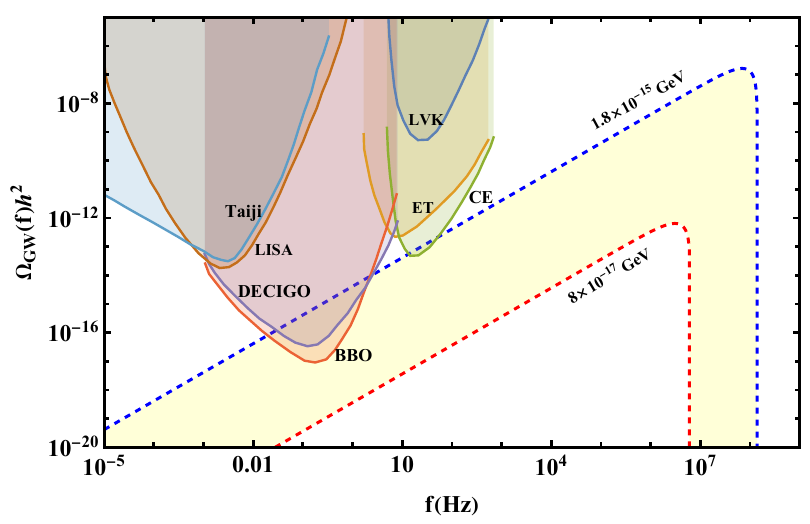}~~~~
\includegraphics[scale=0.515]{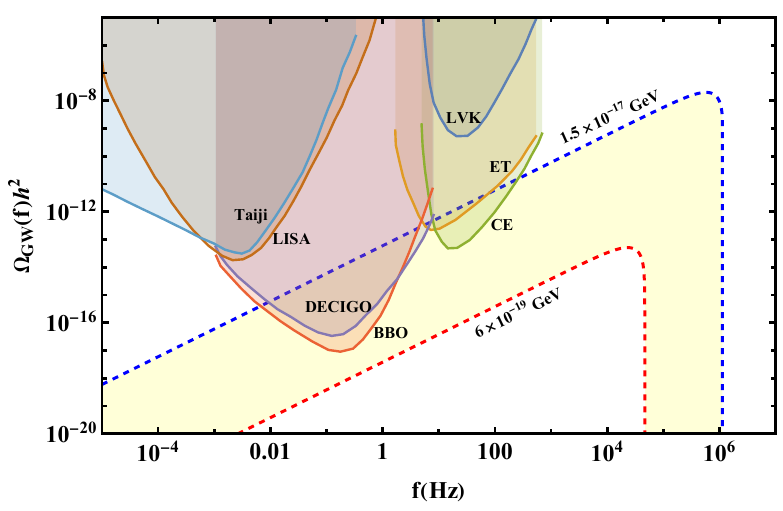}
\caption{Secondary GW spectra for superradiant instability of PBH of mass $4.6\times 10^{-4}M_{\odot}$ (left) and $4.6\times 10^{-2}M_{\odot}$ (right). For both cases the relevant ULB masses have been shown. The relevant sensitivity curves have also been shown.}
\label{bp1sec}
\end{figure}

%%%%%%%%%%%%%%%%%%%%%%%%%%%%%%%%%%%%%%%%%%%%%%%%%%%%%%
As mentioned earlier, the gravitational interactions due to the mid-mass PBHs give rise to negligible SGWB and hence for them we only consider superradiant instability as the main source of the secondary GW. In the left panel of Fig. \ref{bp1sec}, we show the GW spectra due to the superradiant instability of PBH of mass $4.6\times 10^{-4}M_{\odot}$ in presence of ULBs of different masses, varyoing from $1.8\times 10^{-15}\mathrm{~GeV}$ to $8\times 10^{-17}\mathrm{~GeV}$. It can be seen from the figure that the GW spectra for the highest mass ULB is the strongest and vice versa.
Similarly, in the right panel of Fig. \ref{bp1sec} we have show the GW spectra due to the superradiant instability of PBH of mass $0.046M_{\odot}$ for the presence of ULBs of masses varying from $1.5\times 10^{-17}\mathrm{~GeV}$ to $6\times 10^{-19}\mathrm{~GeV}$. Similar to the previous case, here also more massive ULBs correspond to stronger GW signal.
It is to be noted here though for the mid-mass PBHs, some part of the secondary GW spectra is within BBO, DECIGO, CE amd ET, the peaks of these spectra are well beyond the reach of any observatories.
%%%%%%%%%%%%%%%%%%%%%%%%%%%%%%%%%%%%%%%%%%%%%%%%%%%%%%
\subsection{High-mass PBH}
\label{highpbh}
\begin{figure}[t]
\centering
\includegraphics[scale=0.525]{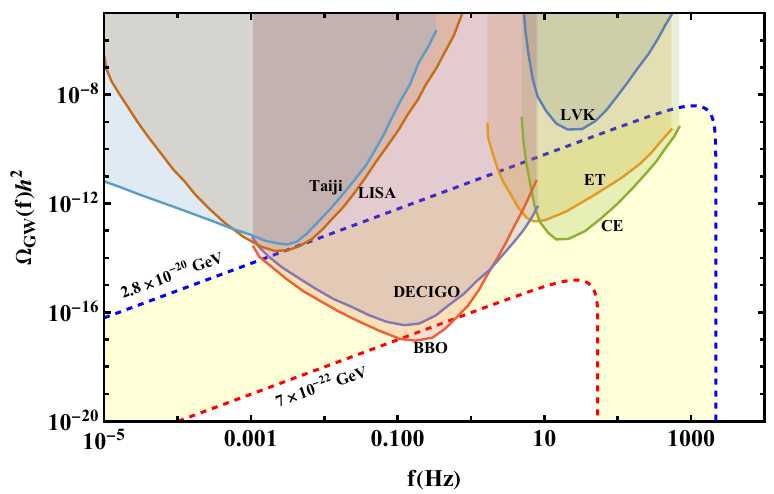}~~~~
\includegraphics[scale=0.258]{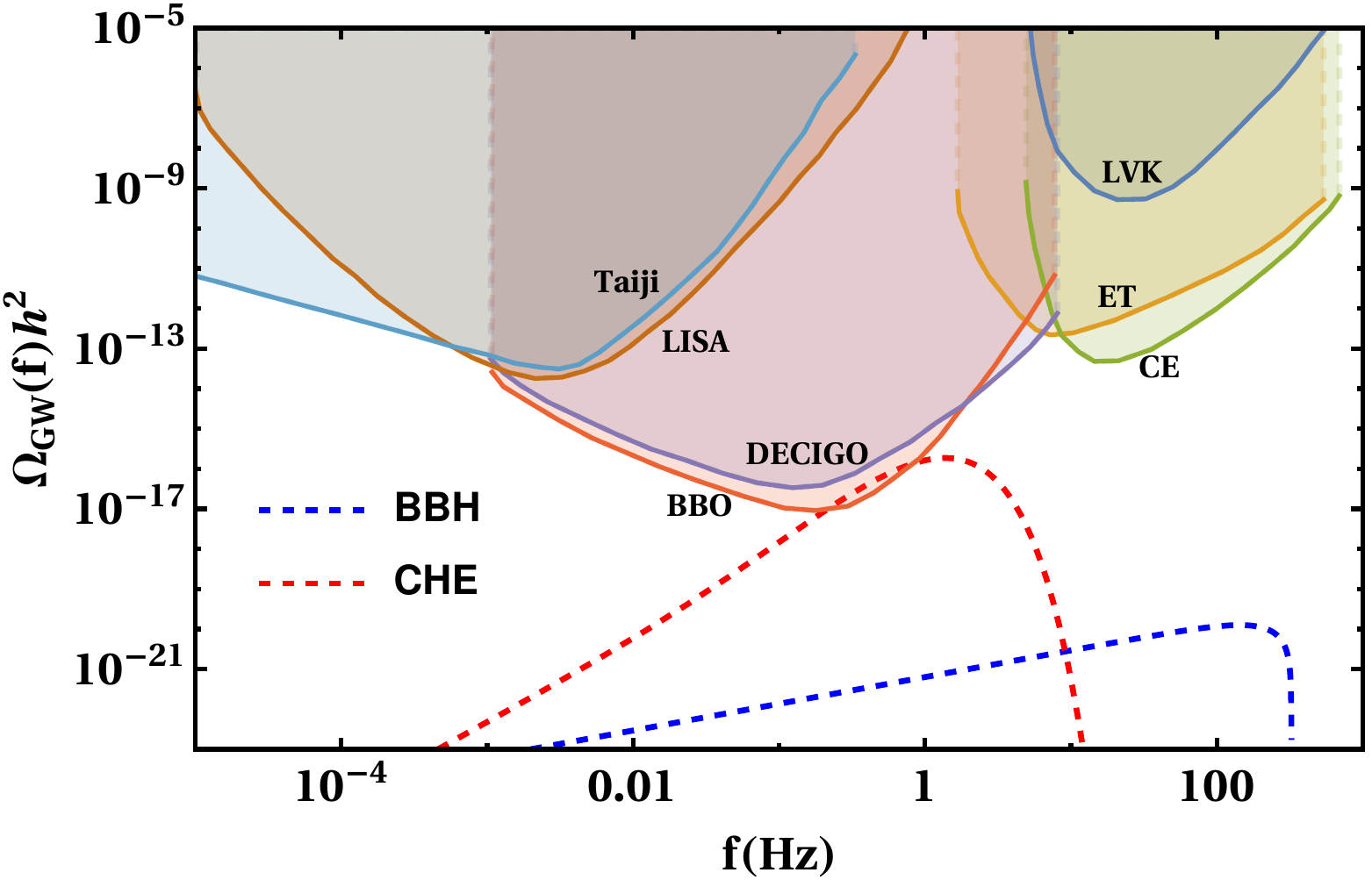}
\caption{(Left) Secondary GW spectra for superradiant instability of PBH of mass $30M_{\odot}$ and for ULB cloud of different masses. (Right) Secondary GW spectra due to the gravitational interactions between the same PBHs. The relevant sensitivity curves have also been shown.}
\label{bp3sec}
\end{figure}

%%%%%%%%%%%%%%%%%%%%%%%%%%%%%%%%%%%%%%%%%%%%%%%%%%%%%%
Contrary to the previous case, for high-mass PBHs, both gravitational interactions, i.e. PBH binary systems (BBH) and close hyperbolic encounters (CHE) and the superradiant instabilities, both give rise to detectable GW spectra. Therefore, in Fig. \ref{bp3sec}, we have shown the GW spectra due to the superradiant instability of PBH of mass $30M_{\odot}$ due to ULBs of masses varying from  $2.8\times 10^{-20}\mathrm{~GeV}$ to $7\times 10^{-22}\mathrm{~GeV}$, as well as the GW spectra due to the PBH BBH and CHE.
In this case superradiant instability creates strong GW signal in the frequency range of the detectors and CHE also generates GW signal just strong enough for BBO, but BBH fails to generate significant GW. However, as it can be seen that for this case, if the mass of the ULB is high enough for the gravitational fine structure constant $\alpha_g>0.01$, the GW due to superradiance completely overshadows the GW due to CHE. Therefore, in the upcoming sections of out study, we ignore the gravitational interactions even in the case of the high-mass PBH.
%%%%%%%%%%%%%%%%%%%%%%%%%%%%%%%%%%%%%%%%%%%%%%%%%%%%%%
%%%%%%%%%%%%%%%%%%%%%%%%%%%%%%%%%%%%%%%%%%%%%%%%%%%%%%
\section{Results}
\label{sec:results}
%%%%%%%%%%%%%%%%%%%%%%%%%%%%%%%%%%%%%%%%%%%%%%%%%%%%%%
%%%%%%%%%%%%%%%%%%%%%%%%%%%%%%%%%%%%%%%%%%%%%%%%%%%%%%
In this article, we take a scenario where PBHs are generated due to the cosmological FOPTs, and these PBHs then interact among themselves gravitationally and may go through superradiant instability in case there are ULBs present. Therefore, we consider the SGWB due to the bubble wall collision during the FOPT (primary spectra) as well as the SGWB due to gravitational interaction and the superradiant instability of the PBHs (secondary spectra) as the direct and indirect consequence of the FOPT. Hence, we call the sum of all these SGWB as the \textit{cumulative spectra}.    
We present our results in two main sections. First we discuss the cumulative SGWB due to a FOPT and then we shift our focus to the significance of the transition temperature from different perspectives.
%%%%%%%%%%%%%%%%%%%%%%%%%%%%%%%%%%%%%%%%%%%%%%%%%%%%%%
\subsection{Cumulative SGWB Spectra}
\label{midpbhresults}
%%%%%%%%%%%%%%%%%%%%%%%%%%%%%%%%%%%%%%%%%%%%%%%%%%%%%%
\begin{figure}[H]
\centering
\includegraphics[scale=0.6]{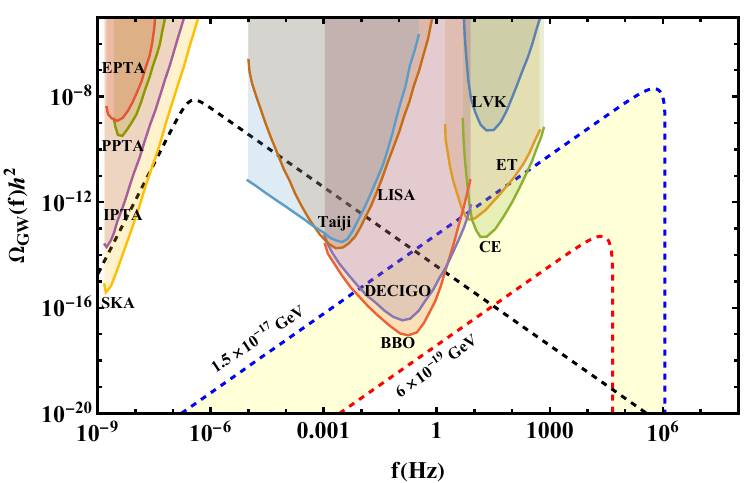}
\includegraphics[scale=0.58]{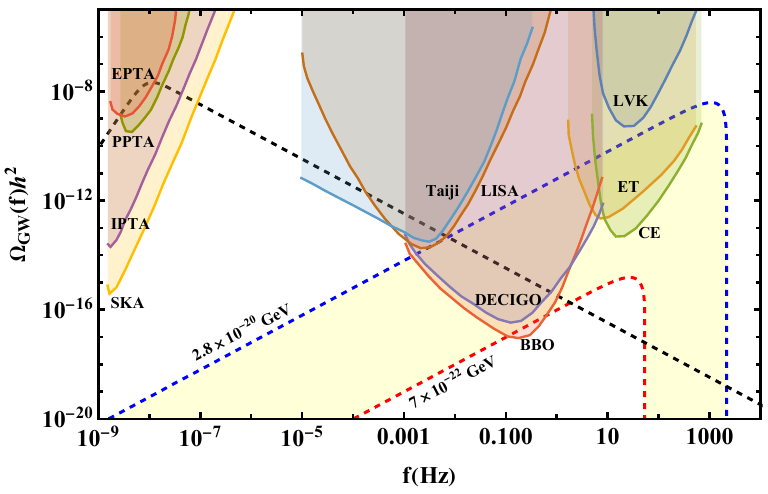}
\caption{(Left) Cumulative spectra due to the FOPT at 1 GeV. (Right) Cumulative spectra due to the FOPT at 0.039 GeV. The relevant sensitivity curves have also been shown.}
\label{bpcum}
\end{figure}
In Fig. \ref{bpcum} we have shown the cumulative spectra due to the FOPTs occurring at 1 GeV and 0.039 GeV respectively. It can be seen from the left panel of the figure that the primary  SGWB due to the FOPT at 1 GeV is marginally within the detection range of SKA and well within the detection range of Taiji, LISA, DECIGO and BBO, whereas the strongest prediction ($m_a=1.5\times 10^{-17}\mathrm{~GeV}$) for the secondary spectra is with in the reach of DECIGO, BBO, CE and ET, however ULBs with $m_a=6\times 10^{-19}\mathrm{~GeV}$ generates very weak spectra which is outside the scope of detection. In the right panel of the figure, it can be seen that the primary spectra due to the FOPT at 0.039 GeV is well within the detection range of the PTA experiments as well as Taiji, LISA, DECIGO and BBO. The secondary spectra for $m_a=2.8\times 10^{-20}\mathrm{~GeV}$ is well within the detection range of DECIGO, BBO, CE, ET. However, the secondary spectra for $m_a=7\times 10^{-22}\mathrm{~GeV}$ is only within the detection range of BBO.
%%%%%%%%%%%%%%%%%%%%%%%%%%%%%%%%%%%%%%%%%%%%%%%%%%%%%%
\subsection{Role of Transition Temperature}
\label{midpbhresults}
%%%%%%%%%%%%%%%%%%%%%%%%%%%%%%%%%%%%%%%%%%%%%%%%%%%%%%
Now we discuss the importance of the transition temperature for the ULB mass and the peak frequency of the primary and secondary spectra. We also present the cumulative spectra for different transitions temperatures in order to gain further insight on the role of the transition temperature.

As mentioned in the previous section, for a particular PBH mass, ULBs with a very narrow range of mass contribute to the secondary spectra formation from superradiant instability as for this study we consider case where $\alpha_g\leq 0.5$. Now, since we that PBH mass depends on the transition temperature, and the ULB mass depends on $\alpha$ and the transition temperature, we can express the relevant ULB mass range as a function of the transition temperature. In this study, we have considered $0.01\leq \alpha_g\leq 0.5$ while presenting our results and therefore, for that range, the ULB mass can be expressed as,
\begin{align}
1.76\times 10^{-19}\left(\dfrac{T}{\mathrm{GeV}}\right)^2\left(\dfrac{g_{*}}{100}\right)^{1/2}\leq m_a \leq 8.8\times 10^{-18}\left(\dfrac{T}{\mathrm{GeV}}\right)^2\left(\dfrac{g_{*}}{100}\right)^{1/2},
\label{maT}
\end{align}
where $T$ is the transtion temperature and $m_a$ is obtained in the units of GeV.
\begin{figure}[t]
\centering
\includegraphics[scale=0.9]{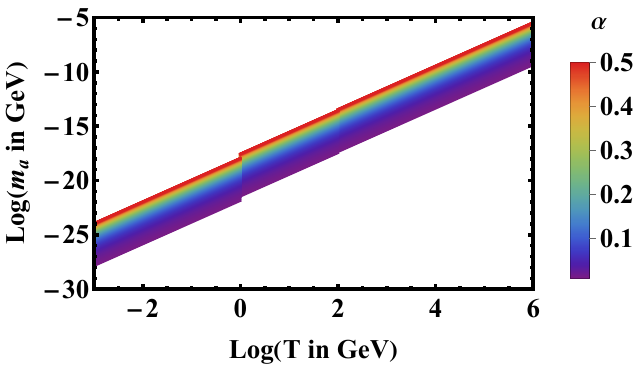}
\caption{Dependence of the relevant value of ULB mass on the transition temperature for the allowd values of gravitaional fine structure constant.}\label{mdepT}
\end{figure}
As expected, it can be seen from Fig. \ref{mdepT} the higher the transition temperature, the higher the mass of the associated ULBs. 

\begin{figure}[t]
\centering
\includegraphics[scale=0.45]{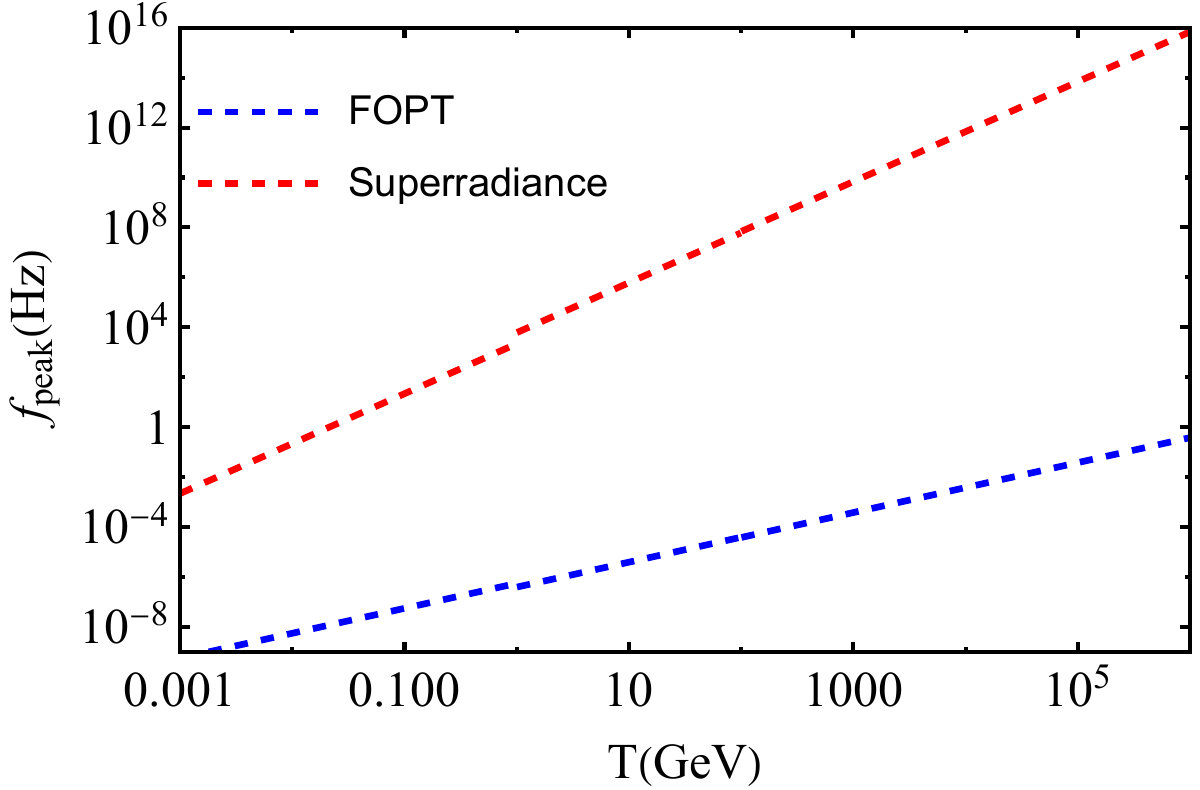}
\caption{The dependence of the peak frequencies of the primary and secondary part of the cumulative spectra on the transition temperature. In this figure we have used $\beta/H=10$ and $\alpha=0.5$.}
\label{peakfreq}
\end{figure}

\begin{figure}[t]
\centering
\includegraphics[scale=0.4]{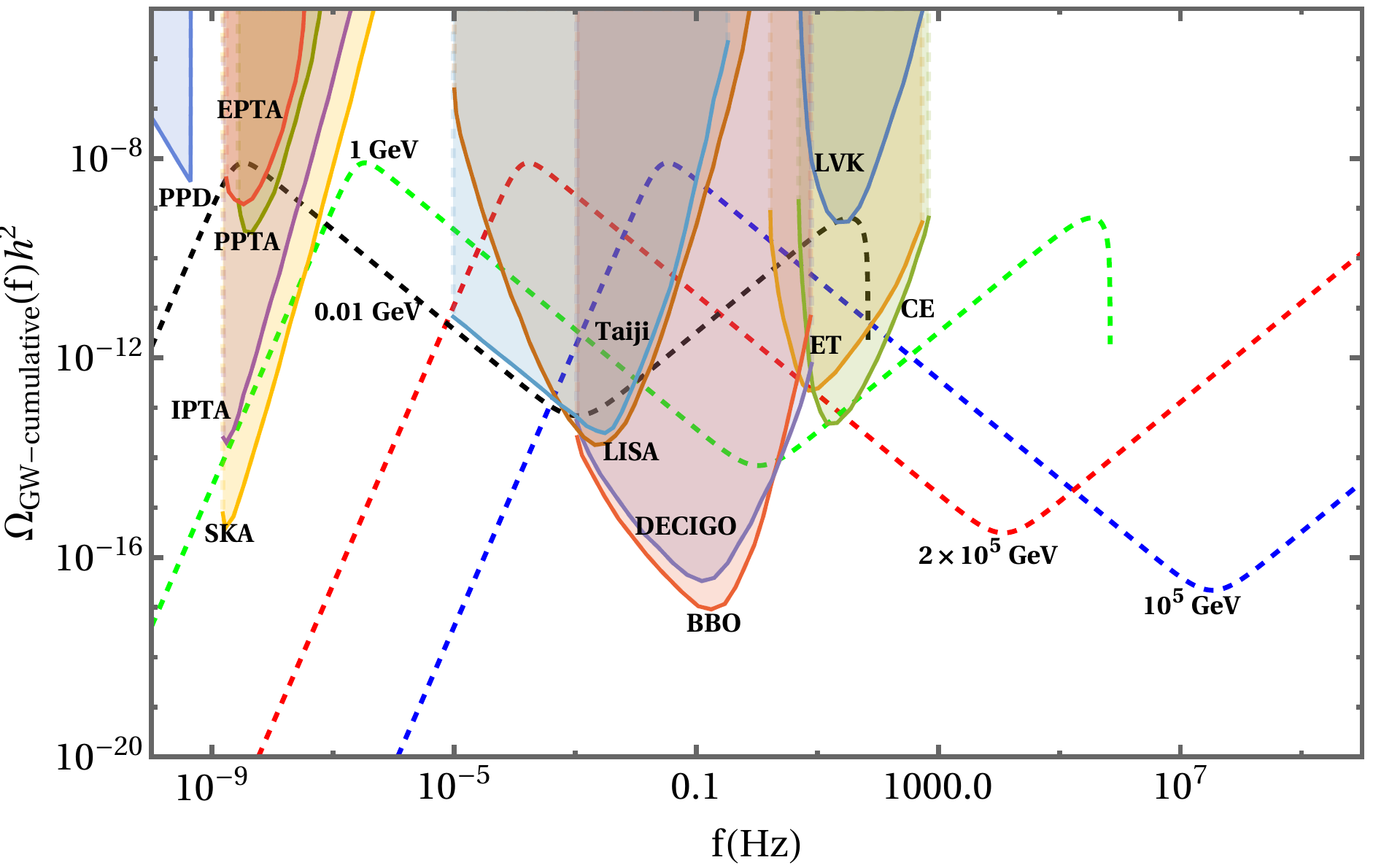}
\caption{The cumulative background for FOPTs at different temperatures, i.e. $0.01\mathrm{~GeV}$ (black dashed), $1\mathrm{~GeV}$ (green dashed), $5\times 10^{2}\mathrm{~GeV}$ (red dashed), and $10^5\mathrm{~GeV}$ (blue dashed). The relevant sensitivity curves have also been shown.}
\label{cumTdep}
\end{figure}

Next we focus on the peak frequency of the cumulative spectra. The primary spectra is due to the FOPT SGWB and its peak frequency is expressed in Eq. \eqref{FOPT_fp}. The secondary spectra in this study is due to the superradiant instability whose peak frequency is given by $f_s = m_a c^2/\pi \hslash$. Since we know that under the conditions of our study, the strongest secondary spectra arises for $\alpha_g=0.5$, therefore, using Eq. \eqref{maT} we express the peak frequency of the secondary spectra as a function of transition temperature,
\begin{align}
f_{\mathrm{peak-suprad}}=6.843\times 10^{3}\left(\dfrac{T}{\mathrm{GeV}}\right)^2\left(\dfrac{g_{*}}{100}\right)^{1/2} \mathrm{~Hz}.
\end{align}
Therefore, in Fig. \ref{peakfreq} we express the dependence of these two peak frequencies on the transition temperature.
It can be seen from the figure that the peak frequency of the primary is much lower than the peak frequency of the secondary throughout the parameter space. Furthermore, the difference between these two peak frequencies is lower at lower temperatures and keeps increasing as the transition temperature increases. To illustrate this further, in the case of a FOPT at $10^{-3}\mathrm{~GeV}$ the difference between the two peak frequencies is $\mathcal{O}(10^6)\mathrm{~Hz}$ whereas for a FOPT occurring at $10^6\mathrm{~GeV}$ the difference is $\mathcal{O}(10^{16})\mathrm{~Hz}$. It is also worth mentioning that for our parameter space, the primary peak frequency is always within the detection range of GW observation experiments. However, most of the secondary peak is outside the scope of detection. However, since these secondary spectra have appreciable amplitude, there is a significant chance that some part of the secondary spectra can be detected. In order to gain deeper insight regarding the possibility of detecting the secondary part, we have shown the cumulative spectra due to FOPTs at different temperatures in Fig.~\ref{cumTdep}.
It can be seen from the figure that FOPT at 0.01 GeV generates the most detectable cumulative background as the primary peak of this can be detected with EPTA, PPTA, IPTA and SKA, while the secondary peak can be detected by ET and CE. For the FOPT at 1 GeV, both the primary and the secondary peaks are outside of the sensitivity curves, however the most of the primary spectrum is within the detection range of SKA, Taiji, LISA, DECIGO, BBO, while some of the secondary peak is marginally within the detection range of DECIGO, BBO and CE. Though the primary peaks from the FOPTs at $5\times 10^{2}\mathrm{~GeV}$ and $10^{5}\mathrm{~GeV}$ are within the detection range of LISA and Taiji, the secondary peaks are entirely out of any possibility of detection. 
%%%%%%%%%%%%%%%%%%%%%%%%%%%%%%%%%%%%%%%%%%%%%%%%%%%%%%
%%%%%%%%%%%%%%%%%%%%%%%%%%%%%%%%%%%%%%%%%%%%%%%%%%%%%%
\section{Summary and Conclusion}
\label{sec:concl}
%%%%%%%%%%%%%%%%%%%%%%%%%%%%%%%%%%%%%%%%%%%%%%%%%%%%%%
%%%%%%%%%%%%%%%%%%%%%%%%%%%%%%%%%%%%%%%%%%%%%%%%%%%%%%
The primary motivation behind this work is to find a method to gain insight into the origin of primordial black holes through gravitational wave backgrounds in the presence of various new physics. In this article, we have considered FOPTs as the generation mechanism of primordial black holes, and we have considered the SGWB due to FOPTs and the SGWB from the superradiant instability of the PBHs. 
In order to extract info from the SGWB, we consider the cumulative SGWB, which is the sum of the primary SGWB created by the FOPT and the secondary SGWB created by the superradiant instability. For illustration, we have taken three benchmark cases of FOPTs in Tab.~\ref{table_FOPT}, i.e., at temperatures $10\mathrm{~GeV}$, $1\mathrm{~GeV}$ and $0.039\mathrm{~GeV}$. The mass and abundance due to these FOPTs are $\{4.6\times10^{-4}M_{\odot}$, $0.01\}$, $\{4.6\times10^{-2}M_{\odot}$, $0.02\}$ and $\{30M_{\odot}$, $0.006\}$. We have shown the primary SGWB for the benchmark cases in Fig.~\ref{FOPT_GW}. In addition, we have considered superradiant instability for the creation of secondary spectra. This is because, for this mass range of PBH, the SGWB due to Hawking evaporation will be of very high frequency and much beyond the scope of being probed. The SGWB due to PBH-PBH or PBH-ABH gravitational interactions in this mass range creates SGWB in the same frequency range as that of superradiant instability, but they are of much lower amplitude. In order to have superradiance, the black holes are to have sufficient angular velocity, but since we consider PBHs created during radiation dominated universe, the initial spin of these PBHs are to be negligible; however, the merger of such PBHs will create spinning PBH, and those PBHs may have superradiant instability (if ULBs of certain mass range exist). In such a case, similar to a normal atom, a gravitational atom is formed around the spinning PBHs, and the ULBs can exist in the ground state or excited state. While going from this excited to this ground state, these ULBs radiate energy as a gravitational wave. It is to be noted here that in order to have very high amplitude SGWB form superradiant instability of a PBH of a certain, there has to exist a ULB of a certain mass. Throughout this article, we have considered a situation where ULBs of various masses exist. To illustrate this further, we have divided our parameter space into two domains, mid-mass PBH ($10^{-6}M_{\odot}-1M_{\odot}$) and high-mass PBH ($1M_{\odot}-10^3M_{\odot}$). We have shown examples of the SGWB due to the PBHs in the mid-mass range in Fig.~\ref{bp1sec}. It can be seen from these figures that in the presence of ULBs of suitable mass superradiance can create detectable SGWB signals. In the case of the high-mass PBHs, we have shown the SGWB due to both gravitational interaction and superradiant instability, and it can be that, in the presence of ULB of suitable mass, the former is negligible when compared to the latter. Hence, in our study, we have only considered the SGWB for the superradiance. It is worth mentioning that apart from the origin of PBH, the absence of these SGWB signals can be used to put bounds on the parameter space of ULB mass.
We show primary, and the secondary SGWB created due to FOPTs as 1 GeV and 0.039 GeV. Along with the primary, in some cases, the secondary part of the SGWB is well within the detection range of various GW detectors. It is worth mentioning that the cumulative SGWB is detectable in the entire frequency range from $\mathcal{O}(1)\mathrm{~nHz}$ to $\mathcal{O}(1)\mathrm{~kHz}$ which gives us a vast window to extract information about the different parts of the SGWB spectra as parts of it can be probed through all proposed GW detectors.
Furthermore, we investigate the role of the transition temperature on the various quantities in our study. It is worth mentioning here that since we are not considering supercooled phase transitions, the transition, nucleation, and percolation temperature are of the same order in this study. Firstly, the mass of the PBH depends on the transition temperature, and since the amplitude of the SGWB is measurable only for a specific mass range of the ULBs corresponding to a specific mass of the PBH, the relevant mass range of the ULB also depends on the transition temperature. To illustrate this further, we have shown said dependence on Fig.~\ref{mdepT}. It can be seen that, for a transition around the temperature 0.01 GeV, the secondary SGWB (spectra due to the superradiance of PBHs) will be the highest if the ULB is of the mass $10^{-23}\mathrm{~GeV}$. Next, we have shown the dependence of the peak frequencies (primary and secondary) on the transition temperature. It can be seen from Fig.~\ref{peakfreq} that both the peak frequencies increase with transition temperature; however, the primary peak frequency is within the detection range for the entire parameter space, whereas the secondary peak frequency is not. It is also worth mentioning that the difference between the two peak frequencies increases with transition temperature. We have shown the complete cumulative spectra for the three benchmark FOPTs we considered in Fig.~\ref{cumTdep}. It can be seen that FOPTs with around or below 1 GeV produce a cumulative signal that can be detected well by the proposed observers. Whereas, for transitions above that temperature, only the primary part of the cumulative signal is within the detection range of the observers. Therefore, this prescription is beneficial to probe heavy PBHs if they originated during a low temperature FOPT. It is also worth mentioning that the absence of these secondary signals can help constrain the parameter space of the mass of the ULBs. 

In conclusion, detecting spectra with these features will help us gain crucial insight into the origin mechanism of PBHs and some other new physics components, such as ULBs and superradiance.

%\appendix

%\section{Appn1}
%\label{appn:1}
%%%%%%%%%%%%%%%%%%%%%%%%%%%%%%%%%%%%%%%%%%%%%%%%%%%%%%
%
%\section{Appn2}
%\label{appn:2}
%%%%%%%%%%%%%%%%%%%%%%%%%%%%%%%%%%%%%%%%%%%%%%%%%%%%%%
 
%\section*{Acknowledgments}
\acknowledgments
IKB thanks Harsha Sankar S H for useful discussions. IKB acknowledges the support by the MHRD, Government of India, under the Prime Minister's Research Fellows (PMRF) Scheme, 2022.

%%%%%%%%%%%%%%%%%%%%%%%%%%%%%%%%%%%%%%%%%%%%%%%%%%%%%%
%%%%%%%%%%%%%%%%%   References   %%%%%%%%%%%%%%%%%%%%%
%%%%%%%%%%%%%%%%%%%%%%%%%%%%%%%%%%%%%%%%%%%%%%%%%%%%%%
\bibliographystyle{JHEP}
\bibliography{PBHGW2_ref.bib}

\providecommand{\href}[2]{#2}\begingroup\raggedright\begin{thebibliography}{10}

\bibitem{Zeldovich:1967lct}
Y.B.~Zel'dovich and I.D.~Novikov, \emph{{The Hypothesis of Cores Retarded
  during Expansion and the Hot Cosmological Model}}, {\emph{Soviet Astron. AJ
  (Engl. Transl. ),} {\bfseries 10} (1967) 602}.

\bibitem{LIGOScientific:2016aoc}
{\scshape LIGO Scientific, Virgo} collaboration, \emph{{Observation of
  Gravitational Waves from a Binary Black Hole Merger}},
  \href{https://doi.org/10.1103/PhysRevLett.116.061102}{\emph{Phys. Rev. Lett.}
  {\bfseries 116} (2016) 061102}
  [\href{https://arxiv.org/abs/1602.03837}{{\ttfamily 1602.03837}}].

\bibitem{Carr:1974nx}
B.J.~Carr and S.W.~Hawking, \emph{{Black holes in the early Universe}},
  \href{https://doi.org/10.1093/mnras/168.2.399}{\emph{Mon. Not. Roy. Astron.
  Soc.} {\bfseries 168} (1974) 399}.

\bibitem{Grillo:1980rt}
A.F.~Grillo, \emph{{Primordial Black Holes and Baryon Production in Grand
  Unified Theories}},
  \href{https://doi.org/10.1016/0370-2693(80)90897-7}{\emph{Phys. Lett. B}
  {\bfseries 94} (1980) 364}.

\bibitem{Khlopov:1980mg}
M.Y.~Khlopov and A.G.~Polnarev, \emph{{PRIMORDIAL BLACK HOLES AS A COSMOLOGICAL
  TEST OF GRAND UNIFICATION}},
  \href{https://doi.org/10.1016/0370-2693(80)90624-3}{\emph{Phys. Lett. B}
  {\bfseries 97} (1980) 383}.

\bibitem{Khlopov:1985jw}
M.~Khlopov, B.A.~Malomed and I.B.~Zeldovich, \emph{{Gravitational instability
  of scalar fields and formation of primordial black holes}}, {\emph{Mon. Not.
  Roy. Astron. Soc.} {\bfseries 215} (1985) 575}.

\bibitem{Carr:1994ar}
B.J.~Carr, J.H.~Gilbert and J.E.~Lidsey, \emph{{Black hole relics and
  inflation: Limits on blue perturbation spectra}},
  \href{https://doi.org/10.1103/PhysRevD.50.4853}{\emph{Phys. Rev. D}
  {\bfseries 50} (1994) 4853}
  [\href{https://arxiv.org/abs/astro-ph/9405027}{{\ttfamily
  astro-ph/9405027}}].

\bibitem{Niemeyer:1997mt}
J.C.~Niemeyer and K.~Jedamzik, \emph{{Near-critical gravitational collapse and
  the initial mass function of primordial black holes}},
  \href{https://doi.org/10.1103/PhysRevLett.80.5481}{\emph{Phys. Rev. Lett.}
  {\bfseries 80} (1998) 5481}
  [\href{https://arxiv.org/abs/astro-ph/9709072}{{\ttfamily
  astro-ph/9709072}}].

\bibitem{Niemeyer:1999ak}
J.C.~Niemeyer and K.~Jedamzik, \emph{{Dynamics of primordial black hole
  formation}}, \href{https://doi.org/10.1103/PhysRevD.59.124013}{\emph{Phys.
  Rev. D} {\bfseries 59} (1999) 124013}
  [\href{https://arxiv.org/abs/astro-ph/9901292}{{\ttfamily
  astro-ph/9901292}}].

\bibitem{Carr:1993aq}
B.J.~Carr and J.E.~Lidsey, \emph{{Primordial black holes and generalized
  constraints on chaotic inflation}},
  \href{https://doi.org/10.1103/PhysRevD.48.543}{\emph{Phys. Rev. D} {\bfseries
  48} (1993) 543}.

\bibitem{Bullock:1996at}
J.S.~Bullock and J.R.~Primack, \emph{{NonGaussian fluctuations and primordial
  black holes from inflation}},
  \href{https://doi.org/10.1103/PhysRevD.55.7423}{\emph{Phys. Rev. D}
  {\bfseries 55} (1997) 7423}
  [\href{https://arxiv.org/abs/astro-ph/9611106}{{\ttfamily
  astro-ph/9611106}}].

\bibitem{Saito:2008em}
R.~Saito, J.~Yokoyama and R.~Nagata, \emph{{Single-field inflation, anomalous
  enhancement of superhorizon fluctuations, and non-Gaussianity in primordial
  black hole formation}},
  \href{https://doi.org/10.1088/1475-7516/2008/06/024}{\emph{JCAP} {\bfseries
  06} (2008) 024} [\href{https://arxiv.org/abs/0804.3470}{{\ttfamily
  0804.3470}}].

\bibitem{Randall:1995dj}
L.~Randall, M.~Soljacic and A.H.~Guth, \emph{{Supernatural inflation: Inflation
  from supersymmetry with no (very) small parameters}},
  \href{https://doi.org/10.1016/0550-3213(96)00174-5}{\emph{Nucl. Phys. B}
  {\bfseries 472} (1996) 377}
  [\href{https://arxiv.org/abs/hep-ph/9512439}{{\ttfamily hep-ph/9512439}}].

\bibitem{Garcia-Bellido:2016dkw}
J.~Garcia-Bellido, M.~Peloso and C.~Unal, \emph{{Gravitational waves at
  interferometer scales and primordial black holes in axion inflation}},
  \href{https://doi.org/10.1088/1475-7516/2016/12/031}{\emph{JCAP} {\bfseries
  12} (2016) 031} [\href{https://arxiv.org/abs/1610.03763}{{\ttfamily
  1610.03763}}].

\bibitem{Braglia:2020eai}
M.~Braglia, D.K.~Hazra, F.~Finelli, G.F.~Smoot, L.~Sriramkumar and
  A.A.~Starobinsky, \emph{{Generating PBHs and small-scale GWs in two-field
  models of inflation}},
  \href{https://doi.org/10.1088/1475-7516/2020/08/001}{\emph{JCAP} {\bfseries
  08} (2020) 001} [\href{https://arxiv.org/abs/2005.02895}{{\ttfamily
  2005.02895}}].

\bibitem{Hawking:1987bn}
S.W.~Hawking, \emph{{Black Holes From Cosmic Strings}},
  \href{https://doi.org/10.1016/0370-2693(89)90206-2}{\emph{Phys. Lett. B}
  {\bfseries 231} (1989) 237}.

\bibitem{Borah:2023iqo}
D.~Borah, S.~Jyoti~Das, R.~Roshan and R.~Samanta, \emph{{Imprint of PBH
  domination on gravitational waves generated by cosmic strings}},
  \href{https://arxiv.org/abs/2304.11844}{{\ttfamily 2304.11844}}.

\bibitem{Crawford:1982yz}
M.~Crawford and D.N.~Schramm, \emph{{Spontaneous Generation of Density
  Perturbations in the Early Universe}},
  \href{https://doi.org/10.1038/298538a0}{\emph{Nature} {\bfseries 298} (1982)
  538}.

\bibitem{Kawana:2021tde}
K.~Kawana and K.-P.~Xie, \emph{{Primordial black holes from a cosmic phase
  transition: The collapse of Fermi-balls}},
  \href{https://doi.org/10.1016/j.physletb.2021.136791}{\emph{Phys. Lett. B}
  {\bfseries 824} (2022) 136791}
  [\href{https://arxiv.org/abs/2106.00111}{{\ttfamily 2106.00111}}].

\bibitem{Baker:2021nyl}
M.J.~Baker, M.~Breitbach, J.~Kopp and L.~Mittnacht, \emph{{Primordial Black
  Holes from First-Order Cosmological Phase Transitions}},
  \href{https://arxiv.org/abs/2105.07481}{{\ttfamily 2105.07481}}.

\bibitem{Huang:2022him}
P.~Huang and K.-P.~Xie, \emph{{Primordial black holes from an electroweak phase
  transition}}, \href{https://doi.org/10.1103/PhysRevD.105.115033}{\emph{Phys.
  Rev. D} {\bfseries 105} (2022) 115033}
  [\href{https://arxiv.org/abs/2201.07243}{{\ttfamily 2201.07243}}].

\bibitem{Kawana:2022olo}
K.~Kawana, T.~Kim and P.~Lu, \emph{{PBH Formation from Overdensities in Delayed
  Vacuum Transitions}},  \href{https://arxiv.org/abs/2212.14037}{{\ttfamily
  2212.14037}}.

\bibitem{Liu:2021svg}
J.~Liu, L.~Bian, R.-G.~Cai, Z.-K.~Guo and S.-J.~Wang, \emph{{Primordial black
  hole production during first-order phase transitions}},
  \href{https://doi.org/10.1103/PhysRevD.105.L021303}{\emph{Phys. Rev. D}
  {\bfseries 105} (2022) L021303}
  [\href{https://arxiv.org/abs/2106.05637}{{\ttfamily 2106.05637}}].

\bibitem{Gouttenoire:2023naa}
Y.~Gouttenoire and T.~Volansky, \emph{{Primordial Black Holes from Supercooled
  Phase Transitions}},  \href{https://arxiv.org/abs/2305.04942}{{\ttfamily
  2305.04942}}.

\bibitem{Lewicki:2023ioy}
M.~Lewicki, P.~Toczek and V.~Vaskonen, \emph{{Primordial black holes from
  strong first-order phase transitions}},
  \href{https://arxiv.org/abs/2305.04924}{{\ttfamily 2305.04924}}.

\bibitem{Matarrese:1993zf}
S.~Matarrese, O.~Pantano and D.~Saez, \emph{{General relativistic dynamics of
  irrotational dust: Cosmological implications}},
  \href{https://doi.org/10.1103/PhysRevLett.72.320}{\emph{Phys. Rev. Lett.}
  {\bfseries 72} (1994) 320}
  [\href{https://arxiv.org/abs/astro-ph/9310036}{{\ttfamily
  astro-ph/9310036}}].

\bibitem{Matarrese:1996pp}
S.~Matarrese and S.~Mollerach, \emph{{The Stochastic gravitational wave
  background produced by nonlinear cosmological perturbations}},  in \emph{{ERE
  - Spanish Relativity Conference}}, 9, 1996
  [\href{https://arxiv.org/abs/astro-ph/9705168}{{\ttfamily
  astro-ph/9705168}}].

\bibitem{Matarrese:1997ay}
S.~Matarrese, S.~Mollerach and M.~Bruni, \emph{{Second order perturbations of
  the Einstein-de Sitter universe}},
  \href{https://doi.org/10.1103/PhysRevD.58.043504}{\emph{Phys. Rev. D}
  {\bfseries 58} (1998) 043504}
  [\href{https://arxiv.org/abs/astro-ph/9707278}{{\ttfamily
  astro-ph/9707278}}].

\bibitem{Khlebnikov:1997di}
S.Y.~Khlebnikov and I.I.~Tkachev, \emph{{Relic gravitational waves produced
  after preheating}},
  \href{https://doi.org/10.1103/PhysRevD.56.653}{\emph{Phys. Rev. D} {\bfseries
  56} (1997) 653} [\href{https://arxiv.org/abs/hep-ph/9701423}{{\ttfamily
  hep-ph/9701423}}].

\bibitem{Easther:2006vd}
R.~Easther, J.T.~Giblin, Jr. and E.A.~Lim, \emph{{Gravitational Wave Production
  At The End Of Inflation}},
  \href{https://doi.org/10.1103/PhysRevLett.99.221301}{\emph{Phys. Rev. Lett.}
  {\bfseries 99} (2007) 221301}
  [\href{https://arxiv.org/abs/astro-ph/0612294}{{\ttfamily
  astro-ph/0612294}}].

\bibitem{Easther:2007vj}
R.~Easther, J.T.~Giblin and E.A.~Lim, \emph{{Gravitational Waves From the End
  of Inflation: Computational Strategies}},
  \href{https://doi.org/10.1103/PhysRevD.77.103519}{\emph{Phys. Rev. D}
  {\bfseries 77} (2008) 103519}
  [\href{https://arxiv.org/abs/0712.2991}{{\ttfamily 0712.2991}}].

\bibitem{Choudhury:2013woa}
S.~Choudhury and A.~Mazumdar, \emph{{Primordial blackholes and gravitational
  waves for an inflection-point model of inflation}},
  \href{https://doi.org/10.1016/j.physletb.2014.04.050}{\emph{Phys. Lett. B}
  {\bfseries 733} (2014) 270}
  [\href{https://arxiv.org/abs/1307.5119}{{\ttfamily 1307.5119}}].

\bibitem{Vilenkin:1981bx}
A.~Vilenkin, \emph{{Gravitational radiation from cosmic strings}},
  \href{https://doi.org/10.1016/0370-2693(81)91144-8}{\emph{Phys. Lett. B}
  {\bfseries 107} (1981) 47}.

\bibitem{Vachaspati:1984gt}
T.~Vachaspati and A.~Vilenkin, \emph{{Gravitational Radiation from Cosmic
  Strings}}, \href{https://doi.org/10.1103/PhysRevD.31.3052}{\emph{Phys. Rev.
  D} {\bfseries 31} (1985) 3052}.

\bibitem{Hindmarsh:1994re}
M.B.~Hindmarsh and T.W.B.~Kibble, \emph{{Cosmic strings}},
  \href{https://doi.org/10.1088/0034-4885/58/5/001}{\emph{Rept. Prog. Phys.}
  {\bfseries 58} (1995) 477}
  [\href{https://arxiv.org/abs/hep-ph/9411342}{{\ttfamily hep-ph/9411342}}].

\bibitem{Witten:1984rs}
E.~Witten, \emph{{Cosmic Separation of Phases}},
  \href{https://doi.org/10.1103/PhysRevD.30.272}{\emph{Phys. Rev. D} {\bfseries
  30} (1984) 272}.

\bibitem{Hogan:1986qda}
C.J.~Hogan, \emph{{Gravitational radiation from cosmological phase
  transitions}}, {\emph{Mon. Not. Roy. Astron. Soc.} {\bfseries 218} (1986)
  629}.

\bibitem{Hawking:1974rv}
S.W.~Hawking, \emph{{Black hole explosions}},
  \href{https://doi.org/10.1038/248030a0}{\emph{Nature} {\bfseries 248} (1974)
  30}.

\bibitem{Ireland:2023avg}
A.~Ireland, S.~Profumo and J.~Scharnhorst, \emph{{Primordial gravitational
  waves from black hole evaporation in standard and nonstandard cosmologies}},
  \href{https://doi.org/10.1103/PhysRevD.107.104021}{\emph{Phys. Rev. D}
  {\bfseries 107} (2023) 104021}
  [\href{https://arxiv.org/abs/2302.10188}{{\ttfamily 2302.10188}}].

\bibitem{Yang:2023aak}
J.~Yang, N.~Xie and F.P.~Huang, \emph{{Implication of nano-Hertz stochastic
  gravitational wave background on ultralight axion particles}},
  \href{https://arxiv.org/abs/2306.17113}{{\ttfamily 2306.17113}}.

\bibitem{Tsukada:2018mbp}
L.~Tsukada, T.~Callister, A.~Matas and P.~Meyers, \emph{{First search for a
  stochastic gravitational-wave background from ultralight bosons}},
  \href{https://doi.org/10.1103/PhysRevD.99.103015}{\emph{Phys. Rev. D}
  {\bfseries 99} (2019) 103015}
  [\href{https://arxiv.org/abs/1812.09622}{{\ttfamily 1812.09622}}].

\bibitem{Berti:2019wnn}
E.~Berti, R.~Brito, C.F.B.~Macedo, G.~Raposo and J.L.~Rosa, \emph{{Ultralight
  boson cloud depletion in binary systems}},
  \href{https://doi.org/10.1103/PhysRevD.99.104039}{\emph{Phys. Rev. D}
  {\bfseries 99} (2019) 104039}
  [\href{https://arxiv.org/abs/1904.03131}{{\ttfamily 1904.03131}}].

\bibitem{Arvanitaki:2010sy}
A.~Arvanitaki and S.~Dubovsky, \emph{{Exploring the String Axiverse with
  Precision Black Hole Physics}},
  \href{https://doi.org/10.1103/PhysRevD.83.044026}{\emph{Phys. Rev. D}
  {\bfseries 83} (2011) 044026}
  [\href{https://arxiv.org/abs/1004.3558}{{\ttfamily 1004.3558}}].

\bibitem{Gehrman:2023esa}
T.C.~Gehrman, B.~Shams Es~Haghi, K.~Sinha and T.~Xu, \emph{{The Primordial
  Black Holes that Disappeared: Connections to Dark Matter and MHz-GHz
  Gravitational Waves}},  \href{https://arxiv.org/abs/2304.09194}{{\ttfamily
  2304.09194}}.

\bibitem{Franciolini:2022htd}
G.~Franciolini, A.~Maharana and F.~Muia, \emph{{Hunt for light primordial black
  hole dark matter with ultrahigh-frequency gravitational waves}},
  \href{https://doi.org/10.1103/PhysRevD.106.103520}{\emph{Phys. Rev. D}
  {\bfseries 106} (2022) 103520}
  [\href{https://arxiv.org/abs/2205.02153}{{\ttfamily 2205.02153}}].

\bibitem{Acuna:2023bkm}
J.T.~Acu\~na and P.-Y.~Tseng, \emph{{Probing primordial black holes from a
  first order phase transition through pulsar timing and gravitational wave
  signals}},  \href{https://arxiv.org/abs/2304.10084}{{\ttfamily 2304.10084}}.

\bibitem{Gehrman:2022imk}
T.C.~Gehrman, B.~Shams Es~Haghi, K.~Sinha and T.~Xu, \emph{{Baryogenesis,
  primordial black holes and MHz\textendash{}GHz gravitational waves}},
  \href{https://doi.org/10.1088/1475-7516/2023/02/062}{\emph{JCAP} {\bfseries
  02} (2023) 062} [\href{https://arxiv.org/abs/2211.08431}{{\ttfamily
  2211.08431}}].

\bibitem{Mandic:2016lcn}
V.~Mandic, S.~Bird and I.~Cholis, \emph{{Stochastic Gravitational-Wave
  Background due to Primordial Binary Black Hole Mergers}},
  \href{https://doi.org/10.1103/PhysRevLett.117.201102}{\emph{Phys. Rev. Lett.}
  {\bfseries 117} (2016) 201102}
  [\href{https://arxiv.org/abs/1608.06699}{{\ttfamily 1608.06699}}].

\bibitem{Chen:2018rzo}
Z.-C.~Chen, F.~Huang and Q.-G.~Huang, \emph{{Stochastic Gravitational-wave
  Background from Binary Black Holes and Binary Neutron Stars and Implications
  for LISA}}, \href{https://doi.org/10.3847/1538-4357/aaf581}{\emph{Astrophys.
  J.} {\bfseries 871} (2019) 97}
  [\href{https://arxiv.org/abs/1809.10360}{{\ttfamily 1809.10360}}].

\bibitem{Sugiyama:2020roc}
S.~Sugiyama, V.~Takhistov, E.~Vitagliano, A.~Kusenko, M.~Sasaki and M.~Takada,
  \emph{{Testing Stochastic Gravitational Wave Signals from Primordial Black
  Holes with Optical Telescopes}},
  \href{https://doi.org/10.1016/j.physletb.2021.136097}{\emph{Phys. Lett. B}
  {\bfseries 814} (2021) 136097}
  [\href{https://arxiv.org/abs/2010.02189}{{\ttfamily 2010.02189}}].

\bibitem{Garcia-Bellido:2021jlq}
J.~Garc\'\i{}a-Bellido, S.~Jaraba and S.~Kuroyanagi, \emph{{The stochastic
  gravitational wave background from close hyperbolic encounters of primordial
  black holes in dense clusters}},
  \href{https://doi.org/10.1016/j.dark.2022.101009}{\emph{Phys. Dark Univ.}
  {\bfseries 36} (2022) 101009}
  [\href{https://arxiv.org/abs/2109.11376}{{\ttfamily 2109.11376}}].

\bibitem{Cui:2021hlu}
W.~Cui, F.~Huang, J.~Shu and Y.~Zhao, \emph{{Stochastic~gravitational~wave
  background from PBH-ABH mergers *}},
  \href{https://doi.org/10.1088/1674-1137/ac4cab}{\emph{Chin. Phys. C}
  {\bfseries 46} (2022) 055103}
  [\href{https://arxiv.org/abs/2108.04279}{{\ttfamily 2108.04279}}].

\bibitem{Papanikolaou:2022chm}
T.~Papanikolaou, \emph{{Gravitational waves induced from primordial black hole
  fluctuations: the~effect of an extended mass function}},
  \href{https://doi.org/10.1088/1475-7516/2022/10/089}{\emph{JCAP} {\bfseries
  10} (2022) 089} [\href{https://arxiv.org/abs/2207.11041}{{\ttfamily
  2207.11041}}].

\bibitem{Xie:2023cwi}
K.-P.~Xie, \emph{{Pinning down the primordial black hole formation mechanism
  with gamma-rays and gravitational waves}},
  \href{https://doi.org/10.1088/1475-7516/2023/06/008}{\emph{JCAP} {\bfseries
  06} (2023) 008} [\href{https://arxiv.org/abs/2301.02352}{{\ttfamily
  2301.02352}}].

\bibitem{Barman:2022pdo}
B.~Barman, D.~Borah, S.~Jyoti~Das and R.~Roshan, \emph{{Gravitational wave
  signatures of a PBH-generated baryon-dark matter coincidence}},
  \href{https://doi.org/10.1103/PhysRevD.107.095002}{\emph{Phys. Rev. D}
  {\bfseries 107} (2023) 095002}
  [\href{https://arxiv.org/abs/2212.00052}{{\ttfamily 2212.00052}}].

\bibitem{Agashe:2022jgk}
K.~Agashe, J.H.~Chang, S.J.~Clark, B.~Dutta, Y.~Tsai and T.~Xu,
  \emph{{Correlating gravitational wave and gamma-ray signals from primordial
  black holes}}, \href{https://doi.org/10.1103/PhysRevD.105.123009}{\emph{Phys.
  Rev. D} {\bfseries 105} (2022) 123009}
  [\href{https://arxiv.org/abs/2202.04653}{{\ttfamily 2202.04653}}].

\bibitem{Datta:2023xpr}
S.~Datta, \emph{{Explaining PTA Data with Inflationary GWs in a PBH-Dominated
  Universe}},  \href{https://arxiv.org/abs/2309.14238}{{\ttfamily 2309.14238}}.

\bibitem{Wang:2016ana}
S.~Wang, Y.-F.~Wang, Q.-G.~Huang and T.G.F.~Li, \emph{{Constraints on the
  Primordial Black Hole Abundance from the First Advanced LIGO Observation Run
  Using the Stochastic Gravitational-Wave Background}},
  \href{https://doi.org/10.1103/PhysRevLett.120.191102}{\emph{Phys. Rev. Lett.}
  {\bfseries 120} (2018) 191102}
  [\href{https://arxiv.org/abs/1610.08725}{{\ttfamily 1610.08725}}].

\bibitem{Wang:2019kaf}
S.~Wang, T.~Terada and K.~Kohri, \emph{{Prospective constraints on the
  primordial black hole abundance from the stochastic gravitational-wave
  backgrounds produced by coalescing events and curvature perturbations}},
  \href{https://doi.org/10.1103/PhysRevD.99.103531}{\emph{Phys. Rev. D}
  {\bfseries 99} (2019) 103531}
  [\href{https://arxiv.org/abs/1903.05924}{{\ttfamily 1903.05924}}].

\bibitem{Wang:2021djr}
S.~Wang, V.~Vardanyan and K.~Kohri, \emph{{Probing primordial black holes with
  anisotropies in stochastic gravitational-wave background}},
  \href{https://doi.org/10.1103/PhysRevD.106.123511}{\emph{Phys. Rev. D}
  {\bfseries 106} (2022) 123511}
  [\href{https://arxiv.org/abs/2107.01935}{{\ttfamily 2107.01935}}].

\bibitem{Choudhury:2023rks}
S.~Choudhury, S.~Panda and M.~Sami, \emph{{Quantum loop effects on the power
  spectrum and constraints on primordial black holes}},
  \href{https://doi.org/10.1088/1475-7516/2023/11/066}{\emph{JCAP} {\bfseries
  11} (2023) 066} [\href{https://arxiv.org/abs/2303.06066}{{\ttfamily
  2303.06066}}].

\bibitem{Choudhury:2023jlt}
S.~Choudhury, S.~Panda and M.~Sami, \emph{{PBH formation in EFT of single field
  inflation with sharp transition}},
  \href{https://doi.org/10.1016/j.physletb.2023.138123}{\emph{Phys. Lett. B}
  {\bfseries 845} (2023) 138123}
  [\href{https://arxiv.org/abs/2302.05655}{{\ttfamily 2302.05655}}].

\bibitem{Basilakos:2023xof}
S.~Basilakos, D.V.~Nanopoulos, T.~Papanikolaou, E.N.~Saridakis and C.~Tzerefos,
  \emph{{Gravitational wave signatures of no-scale Supergravity in NANOGrav and
  beyond}},  \href{https://arxiv.org/abs/2307.08601}{{\ttfamily 2307.08601}}.

\bibitem{Carr:2023tpt}
B.~Carr, S.~Clesse, J.~Garcia-Bellido, M.~Hawkins and F.~Kuhnel,
  \emph{{Observational Evidence for Primordial Black Holes: A Positivist
  Perspective}},  \href{https://arxiv.org/abs/2306.03903}{{\ttfamily
  2306.03903}}.

\bibitem{Karmakar:2023hlb}
R.~Karmakar and D.~Maity, \emph{{Superradiant scattering of electromagnetic
  fields from ringing black holes}},
  \href{https://arxiv.org/abs/2310.01548}{{\ttfamily 2310.01548}}.

\bibitem{Saha:2021pqf}
A.K.~Saha and R.~Laha, \emph{{Sensitivities on nonspinning and spinning
  primordial black hole dark matter with global 21-cm troughs}},
  \href{https://doi.org/10.1103/PhysRevD.105.103026}{\emph{Phys. Rev. D}
  {\bfseries 105} (2022) 103026}
  [\href{https://arxiv.org/abs/2112.10794}{{\ttfamily 2112.10794}}].

\bibitem{Arimoto:2021cwc}
M.~Arimoto et~al., \emph{{Gravitational Wave Physics and Astronomy in the
  nascent era}},  \href{https://arxiv.org/abs/2104.02445}{{\ttfamily
  2104.02445}}.

\bibitem{Banerjee:2023brn}
I.K.~Banerjee and U.K.~Dey, \emph{{Probing the origin of primordial black holes
  through novel gravitational wave spectrum}},
  \href{https://doi.org/10.1088/1475-7516/2023/07/024}{\emph{JCAP} {\bfseries
  07} (2023) 024} [\href{https://arxiv.org/abs/2305.07569}{{\ttfamily
  2305.07569}}].

\bibitem{Goodsell:2009xc}
M.~Goodsell, J.~Jaeckel, J.~Redondo and A.~Ringwald, \emph{{Naturally Light
  Hidden Photons in LARGE Volume String Compactifications}},
  \href{https://doi.org/10.1088/1126-6708/2009/11/027}{\emph{JHEP} {\bfseries
  11} (2009) 027} [\href{https://arxiv.org/abs/0909.0515}{{\ttfamily
  0909.0515}}].

\bibitem{Holdom:1985ag}
B.~Holdom, \emph{{Two U(1)'s and Epsilon Charge Shifts}},
  \href{https://doi.org/10.1016/0370-2693(86)91377-8}{\emph{Phys. Lett. B}
  {\bfseries 166} (1986) 196}.

\bibitem{Calza:2023rjt}
M.~Calz\`a, J.a.G.~Rosa and F.~Serrano, \emph{{Primordial black hole
  superradiance and evaporation in the string axiverse}},
  \href{https://arxiv.org/abs/2306.09430}{{\ttfamily 2306.09430}}.

\bibitem{Ferreira:2020fam}
E.G.M.~Ferreira, \emph{{Ultra-light dark matter}},
  \href{https://doi.org/10.1007/s00159-021-00135-6}{\emph{Astron. Astrophys.
  Rev.} {\bfseries 29} (2021) 7}
  [\href{https://arxiv.org/abs/2005.03254}{{\ttfamily 2005.03254}}].

\bibitem{Freese:2023fcr}
K.~Freese and M.W.~Winkler, \emph{{Dark matter and gravitational waves from a
  dark big bang}},
  \href{https://doi.org/10.1103/PhysRevD.107.083522}{\emph{Phys. Rev. D}
  {\bfseries 107} (2023) 083522}
  [\href{https://arxiv.org/abs/2302.11579}{{\ttfamily 2302.11579}}].

\bibitem{Carena:2019une}
M.~Carena, Z.~Liu and Y.~Wang, \emph{{Electroweak phase transition with
  spontaneous Z$_{2}$-breaking}},
  \href{https://doi.org/10.1007/JHEP08(2020)107}{\emph{JHEP} {\bfseries 08}
  (2020) 107} [\href{https://arxiv.org/abs/1911.10206}{{\ttfamily
  1911.10206}}].

\bibitem{Niikura:2017zjd}
H.~Niikura et~al., \emph{{Microlensing constraints on primordial black holes
  with Subaru/HSC Andromeda observations}},
  \href{https://doi.org/10.1038/s41550-019-0723-1}{\emph{Nature Astron.}
  {\bfseries 3} (2019) 524} [\href{https://arxiv.org/abs/1701.02151}{{\ttfamily
  1701.02151}}].

\bibitem{Griest:2013esa}
K.~Griest, A.M.~Cieplak and M.J.~Lehner, \emph{{New Limits on Primordial Black
  Hole Dark Matter from an Analysis of Kepler Source Microlensing Data}},
  \href{https://doi.org/10.1103/PhysRevLett.111.181302}{\emph{Phys. Rev. Lett.}
  {\bfseries 111} (2013) 181302}.

\bibitem{Niikura:2019kqi}
H.~Niikura, M.~Takada, S.~Yokoyama, T.~Sumi and S.~Masaki, \emph{{Constraints
  on Earth-mass primordial black holes from OGLE 5-year microlensing events}},
  \href{https://doi.org/10.1103/PhysRevD.99.083503}{\emph{Phys. Rev. D}
  {\bfseries 99} (2019) 083503}
  [\href{https://arxiv.org/abs/1901.07120}{{\ttfamily 1901.07120}}].

\bibitem{Macho:2000nvd}
{\scshape Macho} collaboration, \emph{{MACHO project limits on black hole dark
  matter in the 1-30 solar mass range}},
  \href{https://doi.org/10.1086/319636}{\emph{Astrophys. J. Lett.} {\bfseries
  550} (2001) L169} [\href{https://arxiv.org/abs/astro-ph/0011506}{{\ttfamily
  astro-ph/0011506}}].

\bibitem{EROS-2:2006ryy}
{\scshape EROS-2} collaboration, \emph{{Limits on the Macho Content of the
  Galactic Halo from the EROS-2 Survey of the Magellanic Clouds}},
  \href{https://doi.org/10.1051/0004-6361:20066017}{\emph{Astron. Astrophys.}
  {\bfseries 469} (2007) 387}
  [\href{https://arxiv.org/abs/astro-ph/0607207}{{\ttfamily
  astro-ph/0607207}}].

\bibitem{Zumalacarregui:2017qqd}
M.~Zumalacarregui and U.~Seljak, \emph{{Limits on stellar-mass compact objects
  as dark matter from gravitational lensing of type Ia supernovae}},
  \href{https://doi.org/10.1103/PhysRevLett.121.141101}{\emph{Phys. Rev. Lett.}
  {\bfseries 121} (2018) 141101}
  [\href{https://arxiv.org/abs/1712.02240}{{\ttfamily 1712.02240}}].

\bibitem{Murgia:2019duy}
R.~Murgia, G.~Scelfo, M.~Viel and A.~Raccanelli,
  \emph{{Lyman-\ensuremath{\alpha} Forest Constraints on Primordial Black Holes
  as Dark Matter}},
  \href{https://doi.org/10.1103/PhysRevLett.123.071102}{\emph{Phys. Rev. Lett.}
  {\bfseries 123} (2019) 071102}
  [\href{https://arxiv.org/abs/1903.10509}{{\ttfamily 1903.10509}}].

\bibitem{Poulin:2017bwe}
V.~Poulin, P.D.~Serpico, F.~Calore, S.~Clesse and K.~Kohri, \emph{{CMB bounds
  on disk-accreting massive primordial black holes}},
  \href{https://doi.org/10.1103/PhysRevD.96.083524}{\emph{Phys. Rev. D}
  {\bfseries 96} (2017) 083524}
  [\href{https://arxiv.org/abs/1707.04206}{{\ttfamily 1707.04206}}].

\bibitem{Kosowsky:1992vn}
A.~Kosowsky and M.S.~Turner, \emph{{Gravitational radiation from colliding
  vacuum bubbles: envelope approximation to many bubble collisions}},
  \href{https://doi.org/10.1103/PhysRevD.47.4372}{\emph{Phys. Rev. D}
  {\bfseries 47} (1993) 4372}
  [\href{https://arxiv.org/abs/astro-ph/9211004}{{\ttfamily
  astro-ph/9211004}}].

\bibitem{Carilli:2004nx}
C.L.~Carilli and S.~Rawlings, \emph{{Science with the Square Kilometer Array:
  Motivation, key science projects, standards and assumptions}},
  \href{https://doi.org/10.1016/j.newar.2004.09.001}{\emph{New Astron. Rev.}
  {\bfseries 48} (2004) 979}
  [\href{https://arxiv.org/abs/astro-ph/0409274}{{\ttfamily
  astro-ph/0409274}}].

\bibitem{Hobbs:2009yy}
G.~Hobbs et~al., \emph{{The international pulsar timing array project: using
  pulsars as a gravitational wave detector}},
  \href{https://doi.org/10.1088/0264-9381/27/8/084013}{\emph{Class. Quant.
  Grav.} {\bfseries 27} (2010) 084013}
  [\href{https://arxiv.org/abs/0911.5206}{{\ttfamily 0911.5206}}].

\bibitem{Lentati:2015qwp}
L.~Lentati et~al., \emph{{European Pulsar Timing Array Limits On An Isotropic
  Stochastic Gravitational-Wave Background}},
  \href{https://doi.org/10.1093/mnras/stv1538}{\emph{Mon. Not. Roy. Astron.
  Soc.} {\bfseries 453} (2015) 2576}
  [\href{https://arxiv.org/abs/1504.03692}{{\ttfamily 1504.03692}}].

\bibitem{Shannon:2015ect}
R.M.~Shannon et~al., \emph{{Gravitational waves from binary supermassive black
  holes missing in pulsar observations}},
  \href{https://doi.org/10.1126/science.aab1910}{\emph{Science} {\bfseries 349}
  (2015) 1522} [\href{https://arxiv.org/abs/1509.07320}{{\ttfamily
  1509.07320}}].

\bibitem{Ruan:2018tsw}
W.-H.~Ruan, Z.-K.~Guo, R.-G.~Cai and Y.-Z.~Zhang, \emph{{Taiji program:
  Gravitational-wave sources}},
  \href{https://doi.org/10.1142/S0217751X2050075X}{\emph{Int. J. Mod. Phys. A}
  {\bfseries 35} (2020) 2050075}
  [\href{https://arxiv.org/abs/1807.09495}{{\ttfamily 1807.09495}}].

\bibitem{LISA:2017pwj}
{\scshape LISA} collaboration, \emph{{Laser Interferometer Space Antenna}},
  \href{https://arxiv.org/abs/1702.00786}{{\ttfamily 1702.00786}}.

\bibitem{Kawamura:2011zz}
S.~Kawamura et~al., \emph{{The Japanese space gravitational wave antenna:
  DECIGO}}, \href{https://doi.org/10.1088/0264-9381/28/9/094011}{\emph{Class.
  Quant. Grav.} {\bfseries 28} (2011) 094011}.

\bibitem{Phinney:2004bbo}
S.P.~et~al., \emph{{The Big Bang Observer: Direct detection of gravitational
  waves from the birth of the Universe to the Present,}}, {\emph{NASA Mission
  Concept Study} (2004) }.

\bibitem{Reitze:2019iox}
D.~Reitze et~al., \emph{{Cosmic Explorer: The U.S. Contribution to
  Gravitational-Wave Astronomy beyond LIGO}}, {\emph{Bull. Am. Astron. Soc.}
  {\bfseries 51} (2019) 035}
  [\href{https://arxiv.org/abs/1907.04833}{{\ttfamily 1907.04833}}].

\bibitem{Punturo:2010zz}
M.~Punturo et~al., \emph{{The Einstein Telescope: A third-generation
  gravitational wave observatory}},
  \href{https://doi.org/10.1088/0264-9381/27/19/194002}{\emph{Class. Quant.
  Grav.} {\bfseries 27} (2010) 194002}.

\bibitem{Somiya:2011np}
{\scshape KAGRA} collaboration, \emph{{Detector configuration of KAGRA: The
  Japanese cryogenic gravitational-wave detector}},
  \href{https://doi.org/10.1088/0264-9381/29/12/124007}{\emph{Class. Quant.
  Grav.} {\bfseries 29} (2012) 124007}
  [\href{https://arxiv.org/abs/1111.7185}{{\ttfamily 1111.7185}}].

\bibitem{LIGOScientific:2014pky}
{\scshape LIGO Scientific} collaboration, \emph{{Advanced LIGO}},
  \href{https://doi.org/10.1088/0264-9381/32/7/074001}{\emph{Class. Quant.
  Grav.} {\bfseries 32} (2015) 074001}
  [\href{https://arxiv.org/abs/1411.4547}{{\ttfamily 1411.4547}}].

\bibitem{Barausse:2009uz}
E.~Barausse and L.~Rezzolla, \emph{{Predicting the direction of the final spin
  from the coalescence of two black holes}},
  \href{https://doi.org/10.1088/0004-637X/704/1/L40}{\emph{Astrophys. J. Lett.}
  {\bfseries 704} (2009) L40}
  [\href{https://arxiv.org/abs/0904.2577}{{\ttfamily 0904.2577}}].

\end{thebibliography}\endgroup
%\include{bib}
%%%%%%%%%%%%%%%%%%%%%%%%%%%%%%%%%%%%%%%%%%%%%%%%%%%%%%

\end{document}